\documentclass[12pt]{article}
\usepackage{epsfig}
\usepackage{pstricks}

\newcommand{\be}{\begin{equation}}
\newcommand{\ee}{\end{equation}}

\def\la{\mathrel{\mathpalette\fun <}}

\def\fun#1#2{\lower3.6pt\vbox{\baselineskip0pt\lineskip.9pt
\ialign{$\mathsurround=0pt#1\hfil##\hfil$\crcr#2\crcr\sim\crcr}}}

\newcommand{\boalpha}{\mbox{\boldmath $\alpha$}}
\newcommand{\bosigma}{\mbox{\boldmath $\sigma$}}
\newcommand{\borho}{\mbox{\boldmath $\rho$}}

\begin{document}
\begin{center}
 {\bf \Large   The first dozen years of the history    \\[2mm]

of ITEP Theoretical Physics Laboratory}

 \vspace{10mm}

 B.L.Ioffe\\[2mm]

{\it A.I.Alikhanov Institute of Theoretical and Experimental Physics (ITEP),}\\[2mm]

{\it B.Cheremushkinskaya 25, 117218, Moscow, Russia}

\end{center}

\vspace{1cm}

\begin{abstract}


 The theoretical investigations at ITEP in the years 1945-1958 are reviewed.
 There are exposed  the most important theoretical results,
 obtained in the following branches of physics: ~1) the theory of
 nuclear reactors on thermal neutrons; ~2) the hydrogen bomb
 project (``Tube'' in USSR and ``Classical Super'' in USA); ~3)
 radiation theory; ~4) low temperature physics; ~5) quantum
 electrodynamics and quantum field theories; ~6) parity violation
 in weak interactions, the theory of $\beta$-decay and other weak
 processes; ~7) strong interaction and nuclear physics. To the
 review are added the English translations of few papers,
 originally published in Russian, but unknown (or almost unknown)
 to Western readers.

\end{abstract}

\newpage





\newpage

\section{How I became a theoretical physicist}

In the war time, when the German forces came close to Moscow, I with my mother were
evacuated from Moscow to Siberia (my father was in the army) and came back  in June  of
1943. I was 17 years old and had finished 9 classes of the middle school. (The middle
school had 10 classes.) The question arose for me: what to do next? To continue learning
in the school made no sense: next year I would be 18 and should go to the army. And such
youths  as I was, having no experience and not adapted to army life, when they appear at
the battle front, as a rule, perished in the first attack. (I new this well, since in
1942 for half a year, I was working in a military hospital.)

At that time, in the summer of 1943,  preparatory classes were open in several Moscow
technical colleges (in USSR they were called Institutes), where the students were taught
engineering specialities. These classes accepted pupils who had finished 9 classes of
middle school. For few months they studied the program of 10-th class and after that
became students of the college. The students of such colleges should  not go to the army:
even in the year 1943 the goverment understood that after the war the country would need
engineers. (The main part of old engineers were killed in the battles.) There were no
preparatory classes in Moscow University, and at energy and aviation colleges the
deadline for acceptions had passed. The only college where it was still possible to enter
preparatory classes was the Moscow Electro-Mechanical Institute of Railway Transport
Engineers (MEMIIT). I became its student and in 1944 finished the 1-st course.

After the 1-st course, in the summer of 1944, students of MEMIIT were sent wood logging
for a month to a village near the Moscow-Volga Canal. We, four boys, were settled in a
small room of a peasant house.
 We were prepared for what would happen and had
taken with us powder against  the bugs, and strew the floor and the walls up to one meter
height. At night, when we woke up and lit a lantern, we found that all the walls above
this line were mowing -- they were occupied by bugs. Then it became clear, that there
were no toilets in the  whole village -- 100 km from Moscow we were in the Middle Ages.

I was not enthusiastic about the prospect of being a railway engineer, I wanted to be a
physicist. I decided to enter the external department of Moscow University,
 where tuition was
proceeding by correspondence. To do that, I needed permission from MEMIIT, but I got that
easily since I had a good reputation there. Moreover, I was given the right to  visit
lectures freely (not obligatory), which was a very rare case. I entered the
correspondence division of the Physical Faculty in autumn of 1944 and successfully passed
all exams in the
 winter session
-- I got the highest marks in all disciplines. In March-April 1945 a new enrollment was
announced for students in a special group at the Physical Faculty. In this group students
of other institutes  were admitted without  any exams. If the student was enlisted in
such group, his former institute had no right to keep him anymore and such person was
free from the conscription to the army. Now it is clear that it was the enrollment for
the atomic project. I stress that  such enrollment proceeded earlier than
 the explosion of the
first atoming  bomb. I tried to enter this special group but was rejected, although I had
an advantage in comparison with other students -- I had passed the first examination
session at the Physical Faculty. David  Kirzhnits (a future corresponding member of the
Russian  Academy of Sciences) was also rejected, although he had a recommendation from
Landau.

It is interesting how he got this recommendation. Kirzhnits was a student at MAI (Moscow
Aviation Institute). The  teacher of physics (a women) noticed his nonordinary abilities.
She knew Landau and told him about  this talented student. Landau invited Kirzhnits for a
conversation. After the conversation Landau said: ``I will write a letter of
recommendation to Predvoditelev -- the dean of the University Physical Faculty.'' He took
a sheet of paper and a pen, sat down and started to think. ``I cannot write ``Dorogoi
Alexander Savvich'' -- said Landau (in Russian ``dorogoi'' is the standard form of
addressing somebody in a letter but it also  means that the addressee is dear to the
author of the letter) -- He is not dear to me. I cannot write ``Uvazhaemyi ..''
(respected): I have no respect for him''. He thought
 a little more, then exclaimed: ``Oh, I will write him:``Dear'' -- in English
 ``dear'' has no definite
meaning. The reason of rejection of Kirzhnits and me was our nationality -- antisemitism
was on the rise.

In the spring I should sit the exams at MEMIIT and at the correspondence division of the
Physical Faculty of the University. I decided to get a bare pass mark in the exams on
railway disciplines in MEMIIT so that it would be easier to leave MEMIIT.  One of the
courses was: ``Fuel, water and lubricant.'' I had not studied it, and looked through the
textbook only for a few hours before the exam.  The examiner asked me: ``There is a
mixture of water with kerosine. How do you separate them?'' I could not give any answer
besides: to leave the mixture to stay for some time, kerosine will be above, the water
below. I got a 5 -- the highest mark.

I passed all exams at the correspondence division of the Physical Faculty with mark 5 and
received very good recommendations from university professors Gradshtein and Modenov.
Starting from autumn  I again began attempts to  transfer from the correspondence
division to the normal Physical Faculty. I did not mention at the University that I am a
student of MEMIIT, otherwise my request would have been rejected.

I had to visit the same deputy dean of the Physical Faculty,  Georgi Petrovich Z., who
had rejected me when I tried to enter  the special group. The result was the same -- he
said: the transfer from the correspondence division to the normal faculty is forbidden.
Then I visited the vice-rector of Moscow University Prof. Spizyn. He said the same, but
added symphatetically: ``You can complain, it can be repeated infinitely.'' ``And where
to complain ?'' -- asked I. ``In the Ministry of Education'' -- was the answer.

I went to the Ministry  of Education, to the Head of University Department. And here I
had luck ! The Head of Department  Zatsepin was a decent person, he said: ``You want to
transfer from the correspondence division to the Physical Faculty ? Please do it.'' ``And
you can sign a letter, that you don't  object ?'' -- asked I. ``Yes, sure'' -- he
answered. The letter was prepared and signed in a few minutes. With this letter I went to
the Physical Faculty  and here I had luck again ! Georgi Petrovich was on vacation and
instead of him was another man. Looking at him with innocent eyes, I said:``Georgi
Petrovich  promised me that, if I bring a letter from the Ministry of Education, saying
that the Ministry  do not object to my transfer from the correspondence division to the
Physical Faculty, then he would put me in the list of students. Here is such letter.''
``I will do the order'' -- said the man. When Georgi Petrovich came back, he could do
nothing: I already was a student. But I had to leave MEMIIT and leave without a scandal:
overwise my documents would be sent  to military committee and I would be called to the
army although the war was over. I succeeded doing this after 7 visits  to the dean and
rector of MEMIIT.

So, I became a physicist. But a lot of time passed till  the hope appeared that I can be
a theorist.  For a  long time I hesitated, and finally decided to take Landau's
``theoretical minimum'' examination. The first exams -- the entrance exam on mathematics,
the exams on mechanics, classical field theory  and the first part of statistics -- I
passed relatively easily. But in the study of quantum mechanics some problems were
unclear to me: they had to be studied by     reading original papers because the
Landau-Lifshitz textbook did not yet exist at that time.

At the 4-th course the students were distributed among the various departments -
cathedras in Russian. I applied to the  cathedra of theoretical  physics and was
accepted. But soon a new order came  and I was transferred to the cathedra ``The
structure of matter''. This was a coded name; in fact it meant the cathedra of nuclear
and elementary particle physics. I was disappointed by such a decision and tried to have
it revoked, but later I understood that again I had luck. The matter was, that at this
cathedra the diploma supervisor could be any person participating in the atomic project,
 while in other cathedras the supervisor had to be a faculty professor.

Kirzhnits  and I wanted to have a supervisor from the Landau  school. We succeeded to get
two phone numbers -- Pomeranchuk's and Kompaneetz. Kirzhnits started calling Pomeranchuk,
tried a few times, but unsuccessful -- Pomeranchuk was absent. Then he called Kompaneetz
and Kompaneetz agreed to be his supervisor. So, what was left for me -- to call
Pomeranchuk. I did many such calls and got the same answer: `` Pomeranchuk is not here.''
Later, when I started to work at ITEP, I understood, what was the matter. The telephone
of which I had the number, was not in Pomeranchuk's office but in the hall, where a
soldier was sitting. Finally, -- oh miracle --the same voice asked Pomeranchuk to the
phone. Probably, at this moment Pomeranchuk just happened to cross the hall or had a
conversation with somebody in the hall. Again, I had luck! I told Pomeranchuk, that I was
a student at the university, that I had 3 courses of the Landau minimum and I was asking
him to be my diploma supervisor. For Pomeranchuk, the fact, that I had passed 3 courses
of the theoretical minimum was in my favour (at that time not so many people passed the
minimum -- about  10 persons) and he invited me to visit him for a final conversation.
The  day when I came was very cold. I had no fur coat nor even a warm coat -- I was in a
short pilot jacket, sent to USSR by the US within the framework of the lend-lease
program. This jacket determined my fate -- Pomeranchuk agreed -- he told me later, that
the jacket played the role of the last straw.

Pomeranchuk helped me very much with the preparation for the exam on quantum mechanics:
he gave me the proofs of those chapters of ``Quantum mechanics'', where the problems were
expounded that were unclear to me. After that I passed quantum mechanics as well as a few
other courses. I began my diploma work, the theme of which was given to me by
Pomeranchuk. But I was not sure whether I would succeed in becoming a theoretical
physicist. Moreover, I had not much hope to achieve it. As was said once of the heroes of
Sinclair  Lewis' book ``Arrowsmith'': ``Not everybody working in science is a scientist.
Only few of them are.'' Even more this refers to theoretical physics: one thing is to
pass exams and quite another is to work creatively. The first time that hope, that I have
chances to be a theoretical physicist appeared at me during the work on the problem put
to me by Pomeranchuk. I clearly remember this moment -- the starry moment in my life.

Pomeranchuk suggested to me to calculate the polarization of slow (resonance) neutrons
scattering on nuclei. This polarization arises due to interference of the nuclear
scattering amplitude with the amplitude  of the neutron magnetic moment interacting with
the nuclear Coulomb field (relativistic effect). The amplitude of interaction of neutron
magnetic field with Coulomb field is purely imaginary. Therefore interference is possible
only if the nuclear  scattering amplitude has an imaginary part. The similar problem was
solved  previously by Schwinger, who considered neutron scattering  at high energies,
where the nuclear  amplitude is caused by diffraction and is pure imaginary. In the
problem considered by Schwinger, the momentum transfers were much greater than the
inverse sizes of the atom and the interaction of the neutron magnetic moment with atomic
electrons could be neglected. Pomeranchuk suggested to me to consider neutron scattering
at small energies  in the domain of resonances, where the nuclear amplitude also has an
imaginary part. Here, however, the momentum transfers are comparable with inverse atomic
sizes. I tried to find the domain, where it would be possible to apply Schwinger's method
and not consider the interaction of the neutron magnetic moment with atomic electrons,
but the results were  inconclusive  and I was in the dumps. And suddenly, sitting in the
University library -- I clearly remember this moment -- I realized, that I can perform
the calculation exactly: it is necessary to take account of the atomic form factor, and
I can do this by the Thomas-Fermi quasi-classical approach. It was my idea, Schwinger did
not  have it ! And there, the first time, appeared the hope that I could become a
theoretical physicist.

Besides this problem, I solved during my diploma practice two other problems which
Pomeranchuk gave me:  the calculation of  photon polarization by bremsstrahlung and the
production of $e^+e^-$ pairs by polarized photons on a nuclei. At that time there was no
Feynman technique, so I used the old, very complicated (Heitler) technique, where  the
states of electrons  with positive and negative energies were treated separately. I did
not published  these calculations because I considered them as a technical exercise.
Later I  came to regret this strongly , because after a few years several groups
published such calculations. (A short resume  of these papers was published much later
\cite{1}.)


\section{The foundation of ITEP and its program}

ITEP was organized by decree of the USSR government on December 1, 1945. (At that time it
was called Laboratory No. 3 of USSR Academy of Sciences.) Lab. No. 3  was the second,
after Lab. No. 2 (now Kurchatov Institute), organized within the framework of the Soviet
Atomic Project. In fact Lab. No. 3 did not belong to the Academy  of Sciences, but to the
First Directorate of the Council of Ministers of the USSR, which was responsible for the
atomic project. The problems, which should be investigated by ITEP were formulated as:
\\-- physical investigation, design and construction  of nuclear uranium reactor
with heavy water as moderator; \\-- physical investigations of thorium-heavy water and
thorium-uranium-heavy water systems for production of uranium -233;
\\--
physical investigations of $\beta$-radioactivity; \\-- physical investigations of
high energy nuclear particles and cosmic rays.\\
ITEP  theoreticians participated in the investigation  of all these problems, but the
field of their interest was much broader. For the first  half of 1946 the  Head of the
ITEP Laboratory of Theoretical Physics was L.D. Landau, after that I.Ya.Pomeranchuk took
this place, but Landau was continuing working in the Theoretical Laboratory  and came to
ITEP regularly every week till 1958.


\section{The theory of nuclear reactors on thermal neutrons}

The theory of nuclear reactors began with a famous paper by Zeldovich and Khariton
\cite{2}, published in 1940. They introduced in the theory the multiplication coefficient
$k$, defined as the ratio of the number of  neutrons produced in the fission to the
number of absorbed  neutrons in an infinite system, when the loss of neutrons  due to
escape from the system can be  neglected. For $k$ was obtained the formula
\be k=\nu'\varphi \theta\varepsilon, \label{1}\ee
\be \nu' =\nu \frac{\sigma_f}{\sigma_f+\sigma_c}, \label{2}\ee where $\nu$ is the number
of neutrons produced in the process of fission by absorption of a thermal neutron,
$\sigma_f$ is the fission cross section, $\sigma_c$ is the absorption cross section
without fission by fissionable material. (All cross sections refer to thermal  neutrons;
only one fissionable isotope is considered.) In (\ref{1}) $\varphi$ is the probability of
a fast (fission)  neutron to reach thermal energies, escaping the resonance absorption in
the course of moderation; $\theta$ is the probability of absorption of a thermal neutron
by a fissionable nucleus  (for  definiteness let us speak about $^{235}U$), but not by
another nucleus; $\varepsilon$ is the probability for a fast neutron to perform
additional fission in the same uranium   slab. (The factor $\varepsilon$ was absent in
\cite{2}, it was added later by G. Flerov.)  The necessary condition for realization of a
chain  reaction is:
\be k > 1\label{3}\ee

It is easy to find the condition for the reactor to be a breeder, i.e. to produce more
fissionable  material than is burned. Let $\nu'$ neutrons be produced in the fission.
Among them one neutron must  be absorbed  by a fissionable nucleus  in order to have a
chain reaction. One neutron should be absorbed by some material ($^{238}U$ or $Th$) in
order to restore the number of fissionable nuclei. So, the necessary condition for
realization of a breeder is:
\be \nu' > 2\label{4}\ee
 This condition is fulfilled for fast neutrons in the cycle
$^{238}U-Pu(\nu'\approx 2.9)$ and for thermal neutrons in the cycle $Th -
^{233}U(\nu'\approx 2.3)$.

The  most complicated problem in calculating the multiplicity coefficient $k$ is the
calculation of $\varphi$ -- the probability  of the neutron to escape resonance
absorption in the  process of moderation. This problem was solved by Gurevich and
Pomeranchuk  in 1943 \cite{3}. The basic points of their discussion are the following.
Consider a separate resonance at low neutron energy (e.g. the first level of $^{238}U$ at
$E_r=6.67$ eV of width $\Gamma=25\cdot 10^{-3}$ eV and peak absorption cross section
$\sigma_0 =22 \cdot 10^3$ barn). Define \cite{3}  ``the width of the dangerous zone''
$\Delta E_r$ by the equality
\be \sigma_a(E_r +\frac{1}{2}\Delta E_r)\rho \bar{l} =1,\label{5}\ee where $\sigma_a(E)$
is the absorption cross section, $E$ is the neutron energy, $\rho$ is the number of
absorbing nuclei per $cm^3$ of the slab, and $\bar{l}$ is the mean path of the neutron in
the slab. Assume that  in the vicinity of the resonance $\sigma_a(E)$ is described by
the Breit-Wigner formula\\
\be \sigma_a(E) =\sigma_a(E_r) \sqrt{\frac{E_r}{E}} (1+x^2)^{-1},~~~x = \frac{2}{\Gamma}
(E-E_r)\label{6}\ee The substitution of (\ref{6}) in (\ref{5}) gives
\be \Delta E_r  =\Gamma\sqrt{a},~~~a = \sigma_a (E_r)\rho\bar{l}\label{7} \ee For the
first, most important level in $^{238}U$ we have $a \gg 1$ and $\Delta E_r \gg \Gamma$.
This means that the main absorption of neutrons proceeds not in the central part of the
resonance, but in the tales of the resonance curve. (Because of the very large cross
section at the center of the resonance, absorption takes place in the periphery of the
slab, strongly shielding its central  part.)
 If we devide the resonance
levels into  two groups: low energy and high energy resonances, then the discussion
presented above is valid for the former group. For the latter group the resonance
absorption is simply proportional to the number of absorbing nuclei. Using the
Breit-Wigner cross section formula  for  the neutron absorption in the resonance domain,
it is possible to show  that for the former group of levels the resonance absorption is
proportional to $\pi r^2/\sqrt{\bar{l}}$, where $r$ is the slab radius. The final formula
for resonance absorption found in \cite{3} is
\be -\ln \varphi =\frac{\lambda_s}{\xi} \cdot \frac{\alpha d^{3/2} + \beta d^2}{a^2 -\pi
d^2/4}\label{8}\ee
\be \alpha =\frac{\pi}{4} 0.975 \sqrt{\pi
\rho}\sum_i\frac{\sqrt{\sigma_{ai}(E^i_r)\Gamma^r_i}}{E^i_r},~~~\beta=\frac{\pi}{4} \rho
\int\limits^{E_f}_{E_{th}} \sigma(E) \frac{dE}{E},\label{9} \ee where $\lambda_s$ is the
scattering length in the moderator, $\xi$ is the mean logarithmic loss of energy in the
moderator, $1/\xi \approx A/2 +1/3$. In the derivation of (8) it was assumed, that the
generation of resonance neutrons in the moderator is homogeneous in space and isotropic
in velocities. The result (8) is valid for small slabs, where $\bar{l} < \lambda_s$.

In the US the resonance absorption of neutrons was considered by E. Wigner, who proposed
the interpolation  formula
\be
 -\ln \varphi =\frac{\lambda_s}{\xi} \frac{Ad +Bd^2}{a^2-\pi d^2/4}\label{10}
\ee In the first stage of atomic projects in the US and USSR, when the $^{238}U$ level
parameters were not well known, the constants in (8) and (10) were determined empirically
by fitting  the data and the fit was satisfactory for both formulae  (8) and (10). But
even in this case Eq. (8) has an advantage compared with (10): 1) It gives the Doppler
broadening, and hence the temperature  dependence could be predicted \footnote{In some
cases (e.g. the Chernobyl accident) the temperature dependence of the multiplication
coefficient  is extremely important.};
 2) the case could be considered, when the slab contains the moderator,
e.g. when the slab has the shape of a tube with moderator inside (such a calculation was
done by A.Rudik at ITEP \cite{4}).

In 1947 Akhiezer and Pomeranchuk wrote the book ``The Introduction in theory of neutron
multiplying system (reactors)'' \cite{5}, where a detailed exposition of nuclear reactor
theory  was presented. Besides the theory of resonance absorption they considered the
determination of reactor critical size, several aspects of the theory of heterogeneous
reactors, reactor kinetics and temperature coefficient, solved exactly the boundary
condition on a flat boundary of vacuum with medium (by solving the Boltzmann equation by
the Hopf-Wiener method). The book was classified and was published only in 2002. The
theory of heterogeneous reactors was based on a suggestion  formulated by Landau: each
slab is characterized  by a constant $\eta$, equal to the ratio of the flux of thermal
neutrons on its surface to the neutron density there. It was the first  book on the
theory  of nuclear reactors in the world literature, the corresponding books in the
USA appeared later.\\

Other members of the ITEP Theoretical Laboratory also made remarkable contributions to
the theory of nuclear reactors: A.Galanin developed the theory of heterogeneous reactors
\cite{6},\cite{7}; A.Galanin and B.Ioffe performed calculations of the $Th-^{233}U$ cycle
in heavy water reactors and demonstrated the principal realization of the breeder in such
device \cite{8}: B.Ioffe and L.Okun studied the deep burning  of fuel in heavy water
reactors (atomic power stations) \cite{9}. In \cite{9} it was  shown (based on a
suggestion by A.I. Alikhanov), that the continuous  unload of slabs from the reactor has
a serious advantage in comparison with simultaneous unload of all slabs -- the burn-up of
fuel increases by up to a factor 1.5, which is extremely important for atomic power
stations.

On the basis of these  theoretical investigations  the first Soviet research reactor on
natural uranium with heavy water moderator  and cooling was built in 1949 in ITEP. The
critical experiments performed at the start-up of the reactor demonstrated very good
agreement of the  theory with experiment. In 1955 the reactor  was modernized and natural
uranium was changed to enriched uranium. At the start-up of this reactor an accident
happened which will be described in the next Section. Research reactors of this type were
built also under scientific guidance by ITEP in China and  Yugoslavia. According to ITEP
physical and engineering design  several heavy water moderated and cooled reactors for
plutonium production were constructed in the USSR.

In 1972  the first atomic power station in Czechoslovakia -- in Yaslovsky Bohunize
(Slovakia) -- was put in operation, using  natural uranium fuel, heavy water moderator
and gas cooling. Scientific guidance  of the project was performed by ITEP, the physical
design was done by Theoretical Laboratory members Ioffe and Iljichev. The critical
experiment performed at the start-up and the  series  of experiments done after the
startup demonstrated excellent agreement of theory with the data \cite{10}.

\section{Prevention of a dangerous accident at the ITEP reactor in 1955}

The ITEP heavy water research reactor was reconstructed in 1955.
 Instead of natural uranium, the modernized reactor was to use 2\% enriched
uranium, all uranium slabs were replaced by ring-shaped ones, some construction elements
were to be changed also. As a result the reactor power should increase 4-fold and the
flux of thermal  neutrons by one order of magnitude. I performed the physical design of
the reactor. It was the first reactor, going into operation, for whose physical design I
was completely responsible. (Previously I did only the calculations of the physics for
projects of reactors which were not actually constructed. A.D.Galanin was the person,
responsible for physical design of constructed  reactors, and I was only an executor.)

So, come the day of physical start-up of the reactor, when the reactor should become
critical. The head of ITEP reactor division and also the person responsible  for the
physical start-up of the reactor, S.Ya. Nikitin invited me to be present at the
experiment.

The physical start-up of a heavy water reactor proceeds in the following way. The uranium
slabs are inserted in the reactor, in which there is no moderator -- no heavy water.
Since there is no moderator, no chain reaction is going and there is no neutron flux.
Then heavy water is poured in gradually. At some level of heavy water the reactor becomes
critical, the chain reaction begins -- the reactor starts to work. The critical level of
heavy water -- predicted by the theoretical design -- is the main parameter for the
future reactor  exploitation. The coincidence of its experimental value with the
theoretical  prediction means that the theory is reliable enough and that the future
exploitation  of the reactor can proceed based on theoretical predictions. In case of
contradiction  of the theory with experiment any surprises are possible.

Before the beginning of the experiment Nikitin asked  me what the theoretical prediction
for the critical level and its accuracy was. I said that the critical level was 150 cm
and that the error could not exceed 5 cm. The pouring in of the heavy water  began.
Simultaneously the neutron flux N was measured in several places of the reactor (there
was an artificial neutron source at the bottom of the reactor  vessel). The dependence of
1/N as a function of the moderator level was plotted. Evidently, when criticality is
achieved (N turns to infinity),  the curve 1/N must cross the abscissa. The level became
10 cm below my predicted value, then 5 cm below, but the curve 1/N did not look like
going to the expected point. Nikitin consols me: Sometimes it happens, that the  curve
bends in the last moment. The predicted critical level was reached, but the reactor did
not become critical. The  predicted level was exceeded by 5--8 cm, but the reactor did
not start to work. It was easy to understand the thoughts in the minds of all people
attending by looking at their faces:  ``The first reactor was calculated by Pomeranchuk
and Galanin, and see what is happening when this responsible work is entrusted to young
people''. A few more centimeters of heavy water were added but nothing  happened. At this
point Nikitin gave the command to stop the experiment and reported about what had
happened to Abram Isaakovich Alikhanov.

Alikhanov was strongly discontented -- it was greatly annoying for him. It is possible,
that for a moment he had the same thought in his mind as the experimentalists. However,
he postponed all work, related to the start-up of the reactor to the next day and said to
me:``Please check your calculations once more  and report to me  the results tomorrow''.
The whole evening I checked my calculations but did not find any mistakes. I did not
sleep all night. In the morning I collected all my courage, came to Alikhanov and said:
``I do not see any mistakes in the theoretical calculation. There should be no such large
discrepancy between theory and experiment.'' Abram Isaakovich was impressed  by my words
and ordered:  ``Do not perform the setting up of the reactor operation, let the engineers
search for mistakes''.

So,  two days passed. Then  B.A.Medjibovski came to my office. He was an engineer, who
was working on the reactor control system and had nothing to do with reactor construction
and assembly. He asked me: If uranium slabs were fixed not at their right positions, but
20 centimeters higher, what would be the critical level ? I performed the simple
calculation and answered: ``Just at the point where the trend of the  1/N curve was!''
Medjibovski explained  that he had found a place on the blueprints, where by mistake  the
ends of the slabs could be fixed, very similar to the right one but 20 centimeters
higher. He immediately went to tell his guess to Nikitin. Nikitin called A.P.Shilov, the
senior mechanic , the person, responsible for assembling the reactor. Shilov began  to
shout: ``Nonsense! This could not happen ! Never!'' Then Nikitin ordered to take off the
upper cover of the reactor, said that next day he would measure himself in what position
the uranium slabs were fixed and invited me to assist at that. When I came, the upper
cover of the reactor was taken off, Nikitin was standing above. He was in dark goggles,
gloves and overalls. Probably something was put on under the overall. It must be
mentioned, that it is not safe to stay above the reactor when the upper cover is taken
off. Although the reactor was not working, there was still some neutron flux with
associated radiation. Therefore all people should stay at a large distance from the
reactor. Nikitin took a long stick, put it in the reactor, marked some point on the
stick, took it out and measured its length up to the mark. This  procedure he repeated at
several places of the reactor. Then he announced: ``The slabs are fixed at the wrong
places: 20 centimeter higher, than they should be. I will report this to Alikhanov''. The
reactor was reassembled. If the reactor had been put in operation with such a wrong
assembly, then the upper ends of the uranium slabs would have been above the moderator
level, which would result in a strong increase of radiation  caused by fast neutrons, and
the consequence would have been very undesirable. (It must be recalled, that ITEP is
situated in the city of Moscow and the distance from the reactor to the nearest houses is
only  about a few hundred  meters.)


\section{The  hydrogen bomb  project (``Tube'' in USSR and ``Classical Super''
in USA)}

There are two fusion reactions which can be used for the realization of a hydrogen bomb:
\be D + D \to ~^4He +\gamma +23.8 ~\mbox{MeV}\label{11}\ee
\be D + T \to ~^4He + n + 17.6 ~\mbox{MeV}\label{12}\ee At low energies the cross section
of the latter reaction  is about two orders  of magnitude greater than that of the former
one and the process (\ref{12}) proceeds at lower temperatures, than that of  (\ref{11}).
But tritium is unstable (its half life is 12 years), does not exist in nature and can be
produced only in nuclear  reactors in rather small amounts. In 1941 Fermi  in a
conversation with Teller put forward the idea, how the process (\ref{11}) can be used for
the realization of an H-bomb \cite{11}. Suppose that there is a long tube filled with
liquid deuterium. At one  end of the  tube there is an igniter  -- an atomic bomb. When
the atomic bomb explodes a shock wave propagates along the tube, compressing the
deuterium and strongly increasing its temperature: the thermonuclear reaction starts. It
must be stressed that the  length of the tube  is not restricted. Therefore the power of
the bomb is also unlimited. Teller was encouraged by this idea, and began to work on its
realization with his group. In Soviet Union the corresponding project was presented by
Gurevich, Zeldovich, Khariton and Pomeranchuk on 17.12.1945 \cite{12}. I have doubts,
that the idea of the Soviet project was original -- probably it came from intelligence
service, but I believe that the concrete and detailed calculations performed by the
authors of the Russian project were original (see \cite{13},\cite{14}). The possibility
of realizing the project is determined by energy  balance: if it is positive, i.e. the
energy release by nuclear reactions is greater than the energy loss through the surface
of the tube, than the shock wave will propagate along the tube and the hydrogen bomb
explodes, if not, the reaction will dump. The main source of energy loss from  the system
are the $\gamma$-quanta produced by electron bremsstrahlung. The $\gamma$'s can undergo
Compton scattering on electrons. Since the spectrum of $\gamma$'s produced by
bremsstrahlung is softer than the spectrum of electrons, in Compton scattering
$\gamma$-quanta increase their energy in the mean, unlike the case of usual Compton
effect on electrons at rest. (This process can be called inverse Compton-effect.) As a
result of the inverse Compton effect the loss of energy from the system is  increasing.

 Before 1949 no more calculations  were done in
USSR on the project, which got the jargon name ``Tube'', since there was no atomic  bomb
in the USSR -- the igniter, needed for its realization. In 1949 the calculations resumed.
They were performed by Zeldovich's group at Arzamas-16 (now Sarov). Their goal was to
calculate the energy balance. To do this  it was necessary to know the  comptonization
coefficient, defined as the ratio of the energy, carried by $\gamma$'s outside  the tube
after (multiple) inverse Compton collisions, to the energy of bremsstrahlung. The result
of the calculation was, that the balance is zero. But the accuracy of calculations was
low: the accuracy of  the comptonization coefficient calculation was given by a factor of
1.5 or even 2. (E.g. the Compton cross section integrated  over angles was used, whereas
this cross section significantly  depends on angles.) To increase the accuracy to 10\%
requires much more refined methods: taking account of relativistic effects, anisotropy
etc.  At  that time (in 1951) Pomeranchuk was sent to Arzamas-16 by a high level decision
for a long term visit. Pomeranchuk was overburdened by his stay there. It was the time of
fast progress in quantum electrodynamics and Pomeranchuk wanted to participate in this
work, but this was impossible in Arzamas-16. So, he suggested that he and his group would
do the calculations with higher accuracy, up to 10\%, if he would be allowed to go back
to ITEP. He had reasons for such suggestion: at ITEP we were familiar with the Feynman
technique, which was necessary  for relativistic calculations, and at ITEP there was a
strong mathematical group headed by the high class methematician A.Kronrod, who liked
numerical calculations needed in this problem. The  Pomeranchuk proposal was  accepted,
he returned to Moscow and started to form the group. Unfortunately, I became the only
physicist actively participating in the solution of this problem. My colleague A.Rudik
did not get permission to participate although  he was taking part in all calculations of
reactors, including the ones for plutonium and tritium production. The H-bomb project had
the highest degree of secrecy,  and he was not allowed to this level (why this happened
is an enigma to me even now.) Galanin was entirely  busy by reactor calculations,
Berestetsky participated in the solution of particular  problems, separated from the main
one. So, I was left alone. But our group got strong support for numerical calculations.
M.Keldysh, Head of the Committee for mathematical provision of the Atomic Project, got at
our disposal a group of 40 ladies from Leningrad, who performed  calculations on
electro-mechanical calculators. This group was headed by L.Kantorovich, specialist in
numerical calculations and future the Nobel  prize laureate in economics. I calculated
the inverse Compton scattering cross section on arbitrarily moving electrons and
integrated it over the Maxwell spectrum of electrons at given temperature \cite{15}.
Kronrod invented an  effective method of the numerical calculation: it was the solution
of the Peierls equation along the line of flight of the $\gamma$-ray. The work proceeded
in the following way: the Zeldovich group performed the hydrodynamical calculations and
presented to us the distributions of $\gamma$'s, electrons and heavy particles, we
performed the calculations of the energy loss at each point and returned the results to
them back, they perform new hydrodynamical calculations and so on. After 7 iterations the
process converges. The result for energy balance was negative: if the energy, released in
the nuclear reaction is to be taken as 1, then the energy leaving the tube was equal to
1.2. The system was not working, such H-bomb could not be realized in principle. Later it
became known that Teller had come to the same conclusion for the ``Classical Super''.

Mankind was lucky, or maybe God had been merciful.

Since the energy balance was only slightly negative it was necessary to check the
influence of various small effects which were neglected in the calculation. Among them
was the influence of  polarization of the $\gamma$'s. All calculations had been performed
using the Compton cross section averaged over polarization of initial and final photons.
However, the effect of polarization could  be significant. For example in Thomson
scattering (the low energy limit of Compton scattering), if the initial  photon is
unpolarized, then the final photon has a significant polarization. Also, in the same
limit, if the initial photon is completely polarized, then there arises a significant
azimuthal asymmetry in the final photon distributions. The polarization does not affect
the comptonization coefficient for small systems (less than the $\gamma$'s scattering
length), because the polarization can influence the angular distributions only if there
are two or more collisions. On the other hand, the effect of polarization can be
neglected in diffusion approximation, i.e. for large systems (this statement was proven
in \cite{16}). So, one may expect that the influence of polarization is small also in the
intermediate
case.\\
Photon polarization cannot be accounted for within the framework of classical theory --
by solution of Boltzmann or Peierls equations. The  photon polarization can take only two
values, therefore it  is a quantum object. For this reason it was necessary to consider a
$2\times 2$ density matrix. This  was done in \cite{16}, where the following equation for
the density matrix was obtained:
$$
 \frac{\partial J_{\mu\nu}({\bf k},{\bf r})}{\partial t} +{\bf n}{\bf \nabla}
J_{\mu\nu}({\bf k},{\bf r}) + \frac{J_{\mu\nu}({\bf k},{\bf r})}{l(k)}= \int d{\bf
k}'d{\bf p}_1n({\bf p}_1) W_{\mu\nu,\lambda\sigma}(k',k,p_1,p_2) \times
$$
\be \times J_{\lambda\sigma}({\bf k}',{\bf r}) ~\delta(k' +E_{p_1} -k-E_{p_2}) +
F_{\mu\nu}, \label{13} \ee where $J_{\mu\nu}({\bf k},{\bf r})$ is the relativistic
generalization of the density matrix for $\gamma$'s, ${\bf k}$ is the $\gamma$'s
momentum 3-vector, ${\bf n}={\bf k}/k$, $n(p)$ is the Maxwell distribution of electrons,
$F_{\mu\nu}$ is the bremsstrahlung source of $\gamma$'s:
\be F_{\mu\nu} ({\bf k},{\bf r}) = \frac{1}{2} Q({\bf k},{\bf r})
\delta^{\perp}_{\mu\nu},\label{14} \ee $Q({\bf k},{\bf r})$ represents the total number
of gamma's, emitted per unit of time, $\delta^{\perp}_{\mu\nu}$ is the unit tensor with
nonzero components perpendicular to ${\bf k}$, and $l(k)=1/w(k)$. In the limit \\$k/m_e
\ll 1$ the kernel $W_{\mu\nu\lambda\sigma}(k',k,p_1,p_2)$  is given by:
$$
W_{\mu\nu\lambda\sigma}(k',k,p_1,p_2)=\frac{e^2}{4E_1E_2}\frac{1}{kk'}\left\{
\delta_{\mu\lambda} -\frac{1}{(p_1k')}\biggl [ p_{1\lambda} k'_{\mu}
+p_{1\mu}k'_{\lambda}\biggr ] +\frac{kk'}{(p_1k')^2} p_{1\mu} p_{1\lambda}\right \}
\times
$$
\be \times \left\{ \delta_{\nu \sigma} -\frac{1}{(p_1k')} \biggl [ p_{1\sigma}k'_{\nu}
+p_{1\nu} k_{\sigma}\biggr ] +\frac{kk'}{(p_1k')^2} p_{1\nu} p_{1\sigma}\right \} + (\mu
\to \nu,\lambda \to \sigma).\label{15} \ee

\noindent (The exact expression for $W_{\mu\nu\lambda\sigma}$ was presented in
\cite{16}.)
 Equation (\ref{13}) was solved in
\cite{16} in Thomson limit for the case of a plate of thickness of the order of 2-3
scattering lengths. It was found that taking account of polarization for 2 collisions
decreases the yield of photons by by 1-2\%. That means, that the number of Compton
scatterings increases, if the polarization is taken into account. The conclusion was:
taking account of polarization gives of the order of a few percent in the comptonization
coefficient and, probably, results in its increase. So, this statement strengthens the
general conclusion about the impossibility of realization of such a bomb. From the
theoretical point of view it was the first time that the  Boltzmann  equation was written
and solved for a quantum system.

In connection  with this problem an episode happened which I would like to mention here.
One day, when this work was approaching its end, Pomeranchuk came to the office occupied
by Rudik and myself. He said: ``Soon both of you should  present your doctoral theses.
The subject of Rudik's thesis can be declassified, but yours -- he said to me -- must be
classified''. (I remind the reader, that this happened  in the autumn of 1952, when the
wave of antisemitism was growing higher and higher, the ``doctors affair'' was ahead;
Rudik was Russian, I am a Jew.) I presented as my thesis:  ``The influence of
polarization on $\gamma$-quanta propagation in ionized gas''. Although  the purpose of
the calculation was not mentioned in the thesis, moreover the words ``thermonuclear
reaction'' were not mentioned, it had the highest level of secrecy: top secret, special
folder. The approval of  classified theses at that time could be given at only one place:
the Scientific  Council of the Kurchatov Institute. The Chairman of the Council was
Kurchatov, Vice-Chairman was Artsimovich. After  I had presented my thesis and the
opponents -- I.Tamm and I.Khalatnikov -- had given their positive response, one member of
the Council stood up and said: ``O'key, I have no objection. But I have a question. I do
not understand why this thesis has such high level of secrecy ?'' And Artsimovich, who
chaired the meeting, replied: ``And it is quite good that you do not understand!''

Besides the solution of this problem,  V.Berestetsky, I.Pomeranchuk and myself performed
in 1951 another calculation ordered by Zeldovich: The calculation of heat conduction of
completely  ionized gas at high temperatures \cite{17}. This calculation also required a
fully relativistic  treatment: the Boltzmann equation for electron distribution function
must be written in relativistic form. Besides the temperature gradient, the  external
electric field ${\bf E}$ had to be accounted for. The electric field ${\bf E}$ arises
because of redistribution  of charges in the gas, which happens before the equilibration
of temperatures. The value of ${\bf E}$ is determined from the requirement of the
vanishing of the total electric current.

The main idea of the calculation was based on the known fact\cite{18} that large
perpendicular distances to the collision direction play the dominant role in the
gas-kinetical processes in the case of Coulomb interaction. The importance of large
impact parameters corresponds to small momentum transfers in electron-ion collisions. The
results were obtained after tedious numerical calculations. I will not present here the
whole set of results, but restrict myself to presenting the value of heat conductivity
$\kappa$ in the ultrarelativistic case $(T \gg mc^2$):
\be \kappa =0.25 \frac{1}{L} \biggl ( \frac{T}{e^2}\biggr )^2 c,\label{16a}\ee
\be L=\frac{r_{max}}{r_{min}} \approx 10\label{17a}\ee where $r_{max}$ and $r_{min}$ are
the maximal and minimal impact parameters\footnote{This work was done in 1951 and
classified, declassified in 1972 and published in the Collected Works of Pomeranchuk
\cite{17}.}.

\section{Radiation theory}

In 1939 Pomeranchuk  started the study of radiation, emitted by electrons in magnetic
fields. He demonstrated that because of radiation the energy of cosmic ray electrons
falling on the Earth surface is limited, and in particular if the electrons are moving
vertically in the plane of the magnetic equator, then their energy cannot exceed $4\cdot
10^{17}$ eV \cite{19}. The same idea was exploited in  \cite{20}: there it was shown that
in a betatron the energy of circularly orbiting electrons accelerated by the varying
magnetic field inside the circle is limited by
\be E_0 =mc^2 \sqrt{\frac{3}{2}~ \frac{e}{c}~ \frac{R_0}{r^2_0 H^2} \biggl\vert
\frac{dH}{dt}\biggr \vert},\label{16} \ee where $R_0$ is the radius of the orbit and
$r_0=e^2/mc^2$. The angular distribution of the radiation was analyzed in \cite{21}
 and it was shown that it was concentrated
in a small region of angles relative to the plane of the orbit. The spectrum of the
radiation is almost a constant up to $(E_0/mc^2)^3 \omega_0$, where $\omega_0$ is the
orbiting frequency and the spectrum is represented by equidistant lines. This radiation
-- called synchrotron radiation -- is now intensively exploited in many fields: atomic
and molecular physics, physics of condensed matter, biophysics etc. A specially
interesting possibility arises when the electrons are moving in periodically varying
magnetic field along the orbit (the undulator). When the electrons are concentrated  in
bunches  and the number of electrons per bunch is large enough, then it becomes possible
to construct a ``laser on free electrons'' (the undulator, surrounded be optical
reflector), which can produce laser optical beams with frequencies from infra-red up to
ultra-violet. The creation of free electron lasers opens a wide field of new
possibilities in physics and technology.

There is some analogy of the  origin of synchrotron radiation and the deviation of the
bremsstrahlung spectrum in matter from that given by the Bethe-Heitler formula
(Landau-Pomeranchuk effect). Synchrotron radiation is caused by the  curvature of the
electron trajectory in a magnetic field, resulting in acceleration of the electrons. When
an electron passes through matter it undergoes multiple scattering, which results in
electron acceleration in the direction transverse to its line of flight. This
acceleration induces radiation, which  interferes  with bremsstrahlung  and influences
the bremsstrahlung spectrum. Following Landau and Pomeranchuk, consider the case when the
bremsstrahlung photon energy $\omega$ is much less than the electron energy $E$, $\omega
\ll E$. In this case the classical theory of radiation is applicable.

Let us introduce  the coherence length $l_c$ which, by order of magnitude, determines the
longitudinal distance (along the electron momentum) where the bremsstrahlung photon is
formed:
\be
 l_c =\frac{E^2}{m^2 \omega}.\label{16}
\ee At high electron energies the angles of photon emission are small:
\be
 \theta \sim \frac{m}{E}\label{17}
\ee The mean square of the angle of multiple scattering of an electron travelling the
distance $L$ in amorphous  medium is   given by
\be \langle \vartheta^2 \rangle^2_L = \frac{E^2_sL}{E^2L_R},\label{18} \ee where
$E^2_s\approx 4 \pi m^2/e^2$, and $L_R$ is the radiation length
\be L^{-1}_R =\frac{4Z^2 e^6n}{m^2} \ln (183 Z^{-1/2}),\label{19} \ee (here $n$ is the
number of atoms per cm$^3$ and $Z$ is the atomic number). Let us put $L=l_c$. If $\langle
\vartheta^2 \rangle_{l_c}$ exceeds $\theta^2$, then the influence of multiple scattering
on bremsstrahlung radiation becomes important. In the case of condition  $\langle
\vartheta^2 \rangle \gg \theta^2$, which is equivalent to $\omega \ll E^2/E_0$, $E_0
=m^2L_R/E^2_s$, Landau and Pomeranchuk [22] obtained the formula for bremsstrahlung
intensity, emitted per unit of time by electrons in matter:
\be \frac{dI}{d\omega} =2\sqrt{\frac{\omega m^2 e^2}{\pi E^2 L_R}}\label{20} \ee (In
fact, Landau and Pomeranchuk performed the calculations, correct by order of magnitude.
The formula presented above was obtained by Migdal [23], who succeeded in solving the
problem exactly, see also [24].)

The number of quanta, emitted in the frequency interval  $d \omega$ is given by
\be dN = 2\frac{d\omega}{\sqrt{\omega}}\sqrt{\frac{m^2 e^2}{\pi E^2 L_R}}\label{21} \ee
Since the condition $\omega \ll E^2/E_0$
 is fulfilled at small $\omega$, eq.(\ref{21})
states that there is no infrared catastrophe in matter: the total number of emitted
quanta is finite.


\section{Low temperature physics}

The best known work done at ITEP in low temperature physics is Pomeranchuk's
investigation of the properties of $^3He$ [25]. At low temperature  $^4He$  is a
superfluid, which is caused by the Bose-Einstein statistics  of $^4He$ atoms. $^3He$
nuclei have nuclear spin 1/2, so that $^3He$ gas satisfies Fermi-Dirac statistics and the
arguments which hold  for $^4He$ are not valid for $^3He$. Pomeranchuk's considerations
were the following. At low temperatures, when the wave  lengths of $^3He$ atoms are of
order of their distances, the interaction between two $^3He$ atoms depends on the
orientation of their nuclear  spins and exchange effects are important. As a result of
such effects the antiparallel orientation of two neighboring nuclear spins is preferred.
Indeed, in case of antiparallel orientation  the  coordinate wave function can correspond
to $s$-wave, while in case of parallel orientation (the total  spin equal to 1), it is a
$p$-wave. In the former case the distance between two $^3He$ nuclei is smaller and the
attractive  interaction is stronger.

Let us compare the values of entropy  in the solid (crystal) and liquid phases of $^3He$.
In the solid phase, as was shown by Pomeranchuk, the amplitudes of the zero modes of
oscillations are much  smaller than the distances between the atoms. (The crystal state
may be realized at high pressure). So, the  spins of atoms are uncorrelated and the
entropy is a constant equal to $s_{cr}=\ln 2$.  The entropy in the crystal begins to go
to zero at very low temperatures $T_0$. (Pomeranchuk estimated $T_0$ as $T_0\sim
10^{-7}K$; modern estimates give $T_0 \sim 10^{-3}K$ [26].) In the liquid phase, since
the spins are correlated, the entropy is going to zero proportionally to $T$ when $T$
decreases:
\be
 s_{liq} =(T/T_F)\ln 2\label{22}
\ee (See Fig.1). Therefore, at $T_0 <T<T_F$ the entropy of the solid phase is greater
than the entropy of the liquid phase -- the situation is opposite to the standard one. At
the isothermal  melting the heat is secreted, not absorbed  as usually.

In order to find the pressure--$T$-dependence along the phase transition curve, write the
Clausius-Clapeyron relation
\be \frac{dP}{dT}=\frac{s_{liq}-s_{cr}}{v_{liq}-v_{cr}},\label{23} \ee where $v_{liq}$
and $v_{cr}$ are the volume per particle in the liquid and crystal, respectively.
$v_{liq}-v_{cr}$ is always positive. The dependence  $P(T)$ is shown in Fig.2. (The
dashed line corresponds to polarized $^3He$.) At the point $T_m$, defined by
$s_{liq}(T_m) =s_{cr}(T_m)$, the curve $P(T)$ has a minimum. Experimentally, $T_m=0.32K$,
$P_m=29$ bar. According to Pomeranchuk's theory it is possible to cool $^3He$ down to
very low temperatures. The cooling proceeds in the following way. First reach  the point
$p_m,T_m$ by some or other method. Then adiabatically crystalize  the liquid $^3He$ by
increasing the pressure. This process will go along the dashed line on Fig.1 and solid
line on Fig.2. During the process $^3He$ is first solidified and then again is melted.

\begin{figure}[t!]
\begin{minipage}[t]{0.49\textwidth}
 \centerline{\hspace*{-5mm}\includegraphics[height=45mm]{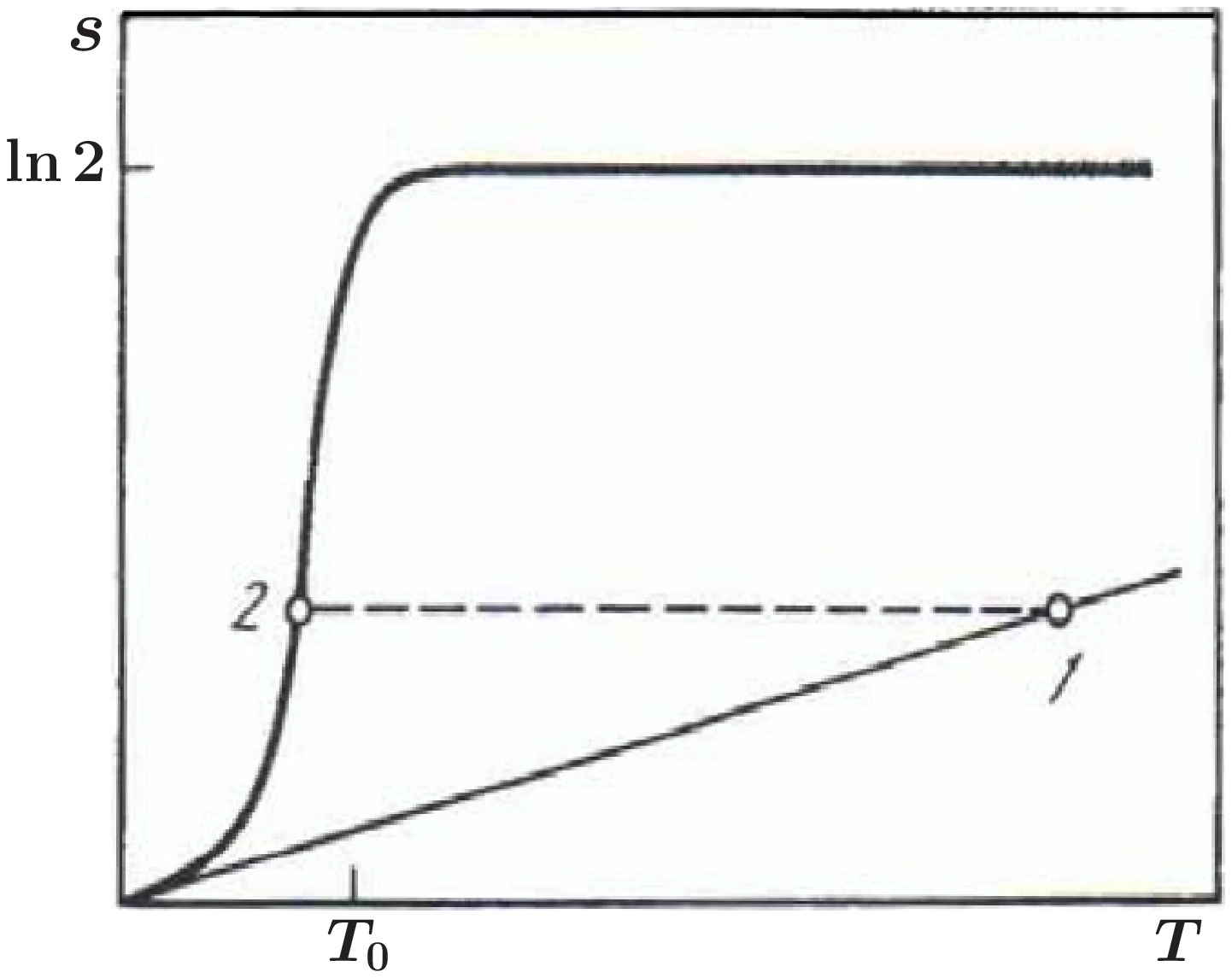}}
  \vspace{-1mm}
   {{\bf Fig. 1.} The entropy dependence on temperature.
    Thick solid line and dashed line corresponds to entropy dependence
    on $T$ in solid and liquid phases at adiabatic cooling
    along the melting curve. The thin solid line corresponds
    to eq.(25).\label{fig:9L}}
\end{minipage}~~~~~~~
\begin{minipage}[t]{0.41\textwidth}
 \centerline{\hspace*{-5mm}\includegraphics[height=45.3mm]{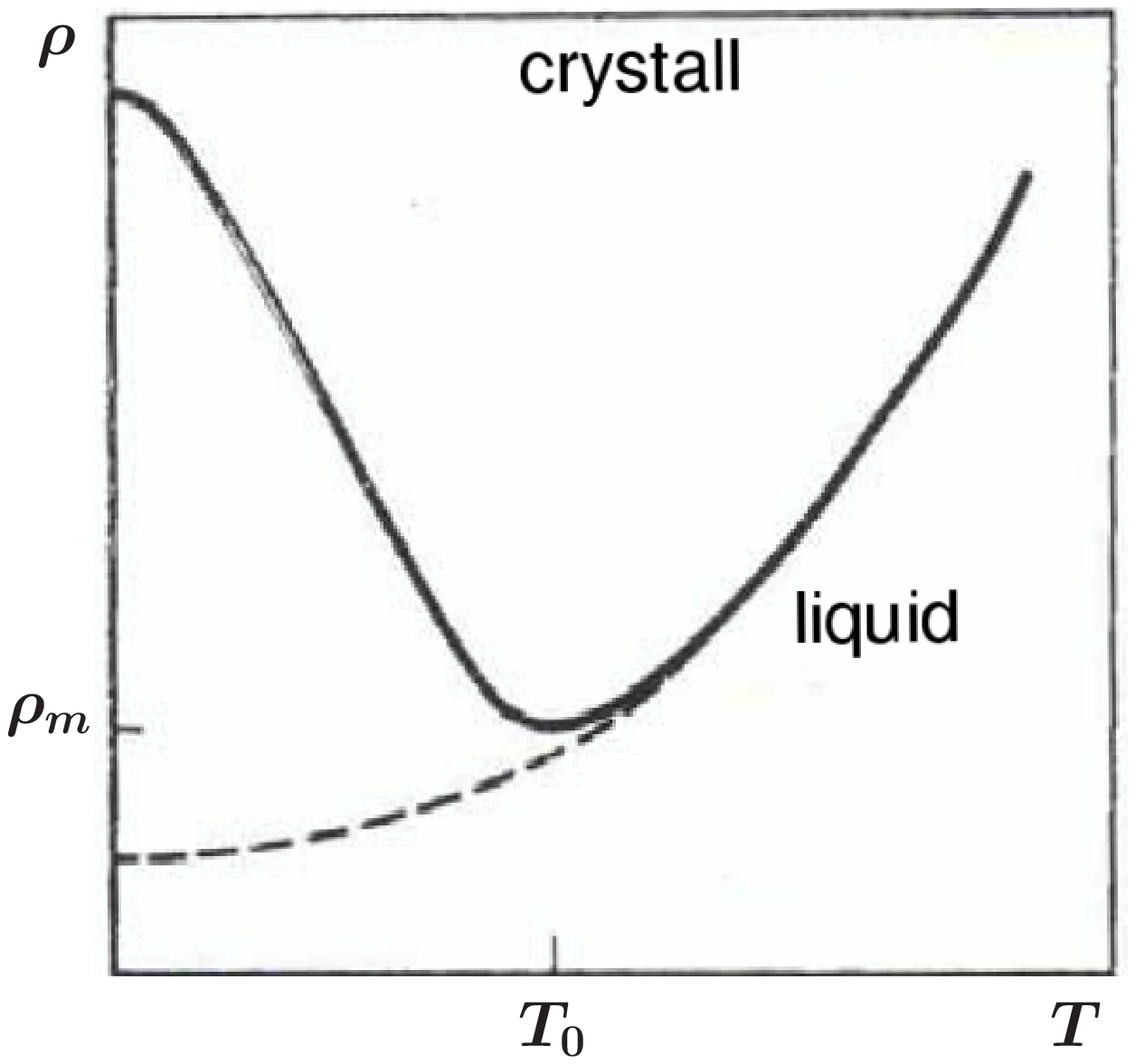}}
  \vspace{-1mm}
   {{\bf Fig. 2.} The dependence of pressure on temperature at pha\-se transition.
    \label{fig:9R}}
\end{minipage}
\end{figure}

In reality, as was found later, Pomeranchuk's reasoning was not completely correct
[26,27]. The wrong assumption was that the distances between $^3He$ atoms are rather
small; in fact in the tunnel exchange three or even more atoms of $^3He$ are
participating. As a consequence, the orbital momentum wave function of the pair of $^3He$
atoms corresponds to $p$-wave and the spin of the pair is equal to 1. But all other
considerations of Pomeranchuk, particularly about entropy behavior in solid and liquid
phases, remain valid also in this case. The only difference is that there are now 3
superfluid phases [27].

Using the Pomeranchuk  method of cooling Osheroff, Richardson and Lee discovered the
superfluidity  of $^3He$. (Nobel prize, 1996) [28]. They found, that  below $3mK$ the
liquid $^3He$ had 3 superfluid phases. All three are magnetic and have anisotropic
behavior.

In paper [25] Pomeranchuk found also the temperature dependences of the specific heat of
$^3He$, its viscosity  and heat conduction. These results were confirmed later by the
Landau theory of Fermi liquids and experiment.

Besides the study of $^3He$, Pomeranchuk (partly together with Landau) investigated the
influence of small admixtures on the properties of liquid helium [29],[30].



\section{Quantum electrodynamics and quantum field theories}

In ITEP we started to study the new approach to quantum electrodynamics -- the Feynman,
Schwinger, Dyson approach -- in 1950. But before this, using the old technique - the
Dirac - Heitler formulation of QED -- several important results were obtained. First,
there was the fundamental theorem, proved by Berestetsky, that the parity of the positron
is opposite to the parity of the electron \cite{32}. Now this statement is trivial, you
can find it in any textbook on QED. But because of its triviality, as a rule there are no
references to the original Berestetsky paper, where the proof was presented for the first
time, and few physicists in the West know by whom and where this theorem has been
formulated. Berestetsky and Pomeranchuk calculated the cross section of $e^+e^-$
annihilation into a $\mu^+\mu^-$ pair \cite{33}. Now this cross section is the standard
theoretical scale and the cross sections of various $e^+e^-\to X$ processes are usually
normalized to it. Pomeranchuk formulated the selection rules for $e^+e^-$ annihilation
into photons \cite{34}. He demonstrated, that the $e^+e^-$ system with total spin $S=1$
cannot annihilate into two photons, and only annihilation into three photons is allowed
for such a system. That means that the  orthopositronium lifetime is much greater (by a
factor of 700) than the lifetime of parapositronium, where two photon annihilation is
allowed. These selection rules were generalized later by the famous Landau-Yang theorem,
which states that two photons with opposite momenta cannot have a total angular momentum
equal to 1. Also mentioned should be the formulation of the electron-positron interaction
up to terms of order to $v^2/c^2$ with account of exchange interaction (Berestetsky,
Landau \cite{35}).

At the beginning of the 50-s only few people in Moscow  studied the new  developments in
quantum field theory: the renormalization of masses and charges, the Feynman diagram
technique  etc. Landau was sceptical about them. Pomeranchuk  tried to persuade him:
``Dau, -- he repeated, -- there are many important and hard problems, which are just for
you''. (Dau, Chuk -- the short names, which  were used very often in Landau circles.)
But Landau declined Chuk's suggestions: ``To solve the problem of infinities  is out of
my possibilities'' he replied.  Two attempts to present talks about Feynman's work at the
Landau seminar were unsuccessful: the speakers were thrown out in the middle of the
seminar. Only the third time the speaker succeeded in finishing his talk, but also not
with full success. Landau preferred to listen about his beloved
 subject -- the alum .
 Landau called me ``snob''. Many times he repeated: ``Boris is a snob''.
  The meaning of his words
was that I did not want to solve real physical problems, but preferred to study refined
physical theories. His words did not influence Pomeranchuk, to whom they were usually
addressed, since we had common views with Pomeranchuk. But, what was the worst, Landau
said the same words to Alikhanov, the ITEP Director. For Alikhanov  Landau was the
unquestionable authority in theoretical physics. Therefore Landau's words could result in
undesirable consequences for me. But luckily Alikhanov had his own opinion in this case.
He knew well that I was performing calculations of nuclear reactors and of his,
Alikhanov's, experimental devices, and that I was no snob by any means. Since Landau did
not allow presentations of new developments in quantum field theory at his seminar,
Pomeranchuk organized his own seminar devoted to quantum field theory. I was the
secretary of this seminar and at the first meeting presented the Dyson paper.

In ITEP we were more interested in meson theories. At that time the only meson found
experimentally was the pion, and it was known that the pion-nucleon coupling constant was
large and perturbation theory was not applicable to the $\pi N$ interaction. (This
statement was demonstrated in \cite{36} by comparison  with experiment of pseudoscalar
meson theory with pseudoscalar and pseudovector couplings.) The idea was to write an
infinite set of coupled equations  for two-, three-, four- ... etc -point Green functions
and perform in this system the mass and charge renormalization. After renormalization
there should be no infinities in the equations. Then to cut the system: to consider, say,
only the first two Green functions  and find the solution neglecting the third, after
that to take account of the third neglecting the fourth and so on. We hoped that such
process would converge. I succeeded in writing such an infinite system of coupled
equations for Green functions using the Schwinger functional equation \cite{37} and
expanding the Green functions, depending on an external source, as a functional series
in this external source \cite{38}. But the equations so obtained were not suitable for
mass and charge renormalization. Then  Galanin, Pomeranchuk and myself \cite{39} used
another form of the Schwinger functional  equations and obtained another infinite set of
coupled equations for many-point Green functions. In this set it was  possible to perform
mass and charge renormalizations. However,  it was impossible to cut the system: the
factor $Z_1$, corresponding to renormalization of vertex functions appeared explicitly
in the equations and in order to calculate it it was necessary to account for the Green
functions beyond the cut. Otherwise the infinities did not cancel. So, this method had to
be abandoned or modified.

In QED Galanin and myself had calculated the first two perturbative terms for the photon
and electron polarization operators and for the vertex function. We found the appearance
of terms $\sim e^2 \ln(\mid p^2\mid /m^2)$ in the first order in $e^2$, where $p$ are the
external momenta -- which we supposed to be of the same order  and $p^2 < 0$. In the
second order appeared terms $\sim (e^2\ln\mid p^2\mid /m^2)^2$. Very instructive for us
was the paper by Edwards \cite{39a}, who constructed  a ladder equation for the vertex
function and found that in the $n$-th order of perturbation theory terms of order
$(e^2\ln\mid p^2\mid /m^2)^n$  persist. Evidently, these terms are the most important at
large $\mid p^2\mid$.
\begin{figure}[t]
\centerline{\includegraphics[width=0.5\textwidth]{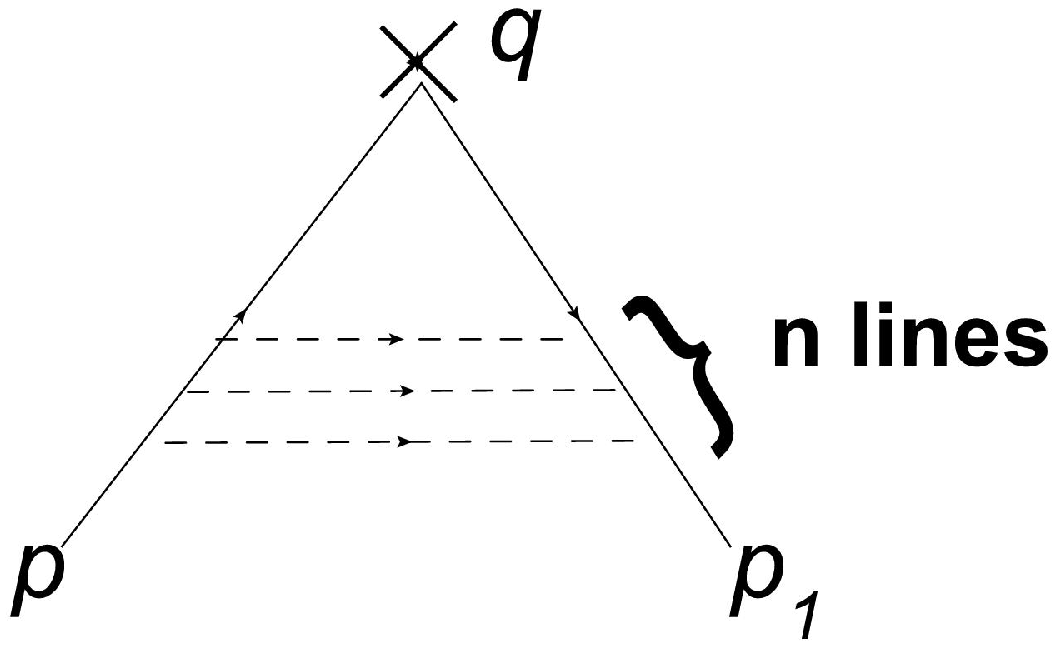}}

\hspace{1.5cm}{\bf Fig. 3.} Ladder approximation for $\bar{e}e\gamma$ vertex function.
Solid and dashed lines
correspond to electron and photon propagators. It is supposed, that $\mid p^2 \mid \sim
\mid p_1^{2}\mid \sim \mid q^2 \mid$. The sum over $n$ should be performed.

\end{figure}


In the 50-s Landau was coming to ITEP every week on Wednesdays, visiting Alikhanov's
experimental seminar. After the seminar he went to the theory room, engaging in
discussions for 1--2 hours. In one such discussion Galanin  and I told him about our
calculations and Edwards results. Next Wednesday Landau showed us his new idea: the
point-like interaction in QED should be changed by a smeared  one, characterized by some
radius $a$ (or, in  momentum space, by the cut-off $\Lambda$). In the calculation with
smeared interaction all the most important terms of order $(e^2\ln\mid p^2\mid
/\Lambda^2)^n$ must be accounted for. In Landau's approach the  equation for the vertex
function corresponds to the diagram of Fig.4.
\begin{figure}[t]
\centerline{\includegraphics[width=0.5\textwidth]{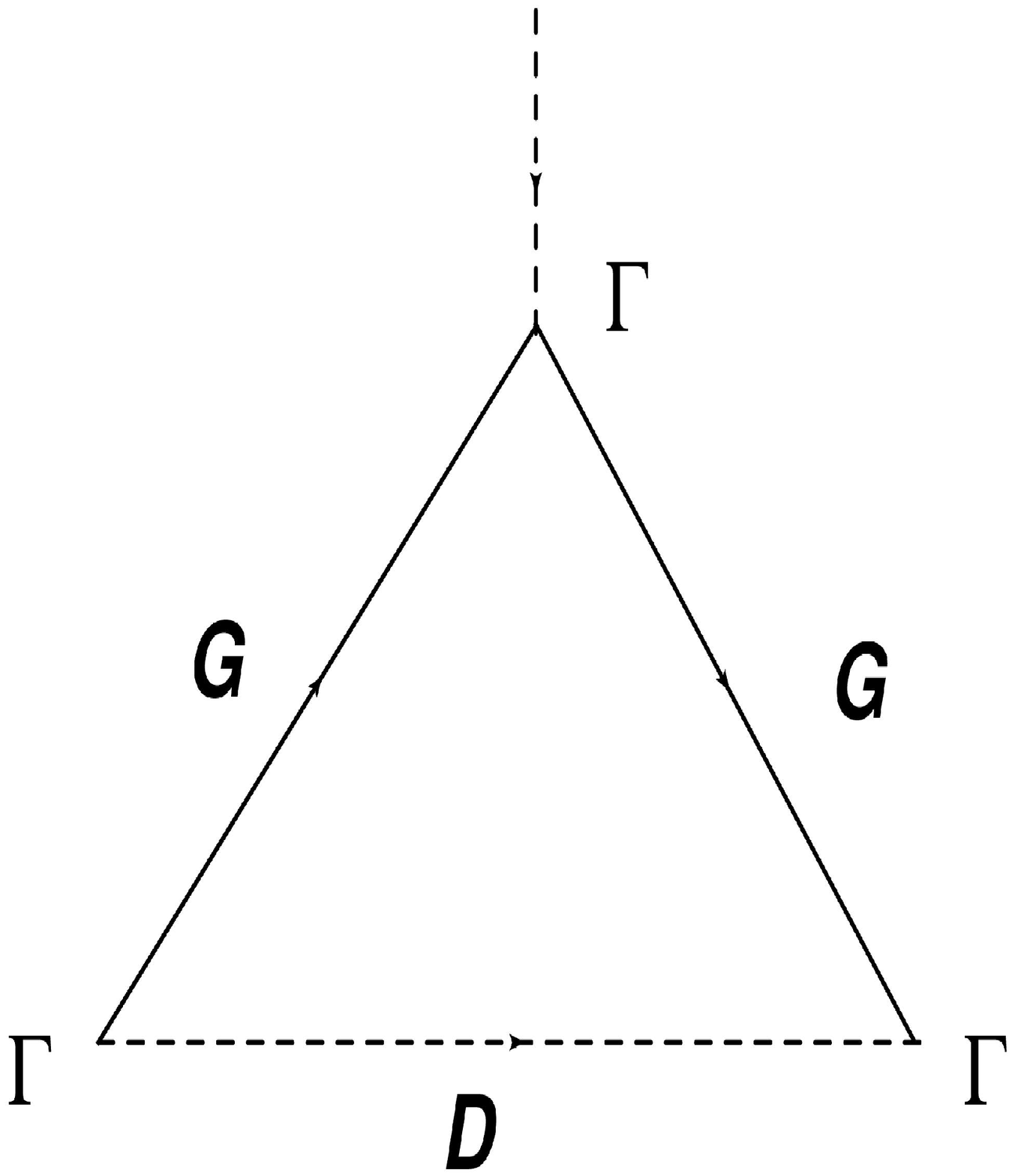}}
\hspace{1cm}{\bf Fig. 4.} Landau equation for vertex function $\Gamma$. $G$ and $D$ --
are exact electron and photon propagators.
\end{figure}
 The equations  for the exact electron  and photon propagators $G$ and $D$ coincide
with the ones proposed by Schwinger and Dyson. Landau expected that after realization of
this program, i.e. after determination of $G$, $D$, and $\Gamma$, it  would be possible
to go to the limit $\Lambda \to \infty$ and all infinities in the theory would disappear.
So, he expected that the theory would be asymptotically free. (This is even more than an
asymptotically free theory, since in asymptotically free theories the infinities do not
disappear: they can only be safely renormalized.) This Landau expectation  can be easily
noticed by looking at the title of the first Landau, Abrikosov and Khalatnikov paper:``
On the removal of infinities in quantum electrodynamics''. The Landau et al. papers were
published as 4 papers in Doklady \cite{40}. The first paper was published before the
whole calculation was finished. Landau did not possess the Feynman technique, therefore
all calculations were done by Abrikosov and Khalatnikov. But the basic ideas  of the
calculations were formulated by Landau: the choice of gauge, the rotation of the
integration contour to the imaginary axis etc.

Here is the suitable place to say few words about Landau. Landau was an  extraordinary
physicist: he knew well the whole of theoretical physics. Not one such physicist existed
in the past and will not exist in future: physics has become too extensive now. There is
a saying attributed to Einstein: God performs integrations in the mind. Landau was not
like that God, but his ability  to do calculations was striking. One example. At the
beginning of the 50-s  Pomeranchuk gave me a problem to solve. I worked about a week and
solved the problem by two methods which, I believed, were equivalent. But the results
were different. I checked the calculations, but found no errors. I told Pomeranchuk about
this discrepancy, and in a few days he performed his own calculations. His results
coincided with mine: the discrepancy still persisted. When  Landau came to ITEP the next
time, Pomeranchuk told him  about this problem. Landau took a stick of chalk and started
to write formulas at the top of the blackboard. After about 20 minutes the result
appeared in the lower part of the blackboard: it coincided  with one of ours. But from
the  derivation it was clear that this result was correct and the other one was wrong.

The main result of Landau, Abrikosov and Khalatnikov's calculations was presented in
terms of an effective electron charge $e^2(\Lambda)$, defined at the scale of cut-off
$\Lambda$. One Wednesday Landau presented to us the result of their calculations:
 \be
  e^2(\Lambda) =\frac{e^2}{1+\frac{e^2}{3\pi}\ln
\frac{\Lambda^2}{m^2}},\label{27} \ee where $e^2$ is the physical charge. The behavior of
$e^2(\Lambda)$ agreed with Landau's expectations: $e^2(\Lambda)$ tends to zero, when
$\Lambda$ tends to infinity. Galanin and  myself decided  to check this result. It was
easy to do it, because Landau, Abrikosov and Khalatnikov had demonstrated that in Landau
gauge, when the photon propagator is transverse, the electron polarization operator and
vertex function do not give terms $\sim e^2\ln(\mid p^2 \mid \Lambda^2)$, and such terms
arise only from the photon polarization operator. We calculated it and found that in
eq.(\ref{27}) there was a mistake of the sign in the denominator. The correct equation
should be
\be e^2(\Lambda) =\frac{e^2}{1-\frac{e^2}{3\pi}\ln \frac{\Lambda^2}{m^2}}\label{28} \ee
The effective electron charge increases with momentum and tends to infinity when
$(e^2/3\pi)\ln (\Lambda^2/m^2)=1$. The correct eq.(\ref{28}), expressing the physical
charge in terms  of the bare charge $e^2(\Lambda)$ defined  at the cut-off, looks like
\be e^2 =\frac{e^2(\Lambda)}{1+\frac{e^2(\Lambda)}{3\pi}\ln
\frac{\Lambda^2}{m^2}}\label{29} \ee So, at arbitrary $e^2(\Lambda)$, when $\Lambda^2 \to
\infty$, the physical charge is going to zero. The theory became contradictory  and the
problem is known now as the problem of vanishing of the physical charge. On the following
Wednesday we told Landau about this error. He checked our  calculation and  agreed with
us. As told by S.Gershtein, who was working at the Landau Theory  Division in  Kapitza's
Institute, Landau came back from ITEP with the words: ``Galanin and Ioffe saved me from
shame !'' (Later Gershtein published this story in his recollections.)

The physical interpretation of eq.(\ref{28}) is the folloing: Landau equations, defined
by contributions of diagrams of Fig.4, are valid at photon and electron virtualities
$p^2$, when
\be 1 -\frac{e^2}{3\pi} \ln \frac{\mid p^2\mid}{m^2} \la e^2\label{30}\ee At higher $\mid
p^2\mid$ the contributions of diagrams, not represented in the form of Fig.4, become
important. A similar situation  arises in Yukawa meson theories as was demonstrated in
\cite{41},\cite{42}. (The latter paper was done within the framework of the approach
[38], when the renormalization was  performed  {\it ab initio},  before the calculation.)

Landau and Pomeranchuk demonstrated, that the trouble arising is not only the defect of
Landau equations, but  is much more general \cite{43}. Their arguments were the
followings. The term 1 in the denominator in eq.(\ref{29}) corresponds to the free term
contribution in the Lagrangian. If this term gives a small contribution in (\ref{29}) at
large $(e^2(\Lambda)/3\pi)\ln(\Lambda^2/m^2)$ and  can be neglected, then there are even
stronger reasons to neglect this term, when the interaction in QED becomes strong, i.e.
the second term in the denominator is very large. In such case we came to the conclusion
that quantum electrodynamics is a contradictory theory. The same conclusion holds  for
meson theories. In QED the contradiction is  nonessential practically, because it arises
at energies unreachable by experiment. In meson theories, since  the coupling constant
$g^2/4\pi$ is large, the contradiction  appears at energies of the order of 1 GeV.
Therefore, such theories must be rejected.

As is well known  now, this problem was solved by the discovery of nonabelian gauge
theories, which have the property of asymptotic freedom.

An important invention in QED was done by Sudakov \cite{44}. He found that in the
electron form factor at high momentum transfer $q=p_2-p_1$, there appear terms
proportional to the square of $\ln q^2$, like $(e^2\ln^2 q^2/m^2)^n$ in $n$-th order of
perturbation theory, and succeeded in summing all of  them. Such terms arise from
diagrams of which an example is shown in Fig.5.
\begin{figure}[t]
\centerline{\includegraphics[width=0.5\textwidth]{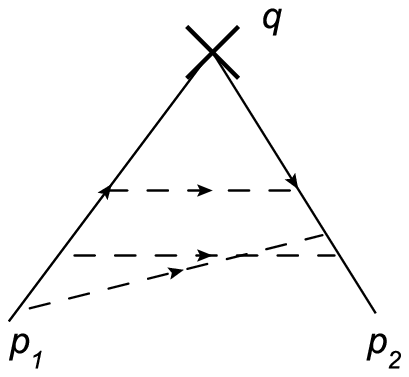}}

\hspace{1.5cm}{\bf Fig. 5.} Example of a diagram where the $\ln^2(q^2/m^2)$ term appear.
\end{figure}
The photon lines connect the initial and final electron propagators. Their number and
their order on both electron  lines is arbitrary and the sum must be performed  over all
such diagrams and over powers $n$ in perturbation  theory. Other diagrams where the
photon line connects the same electron line or diagrams with electron loops do not result
in the appearance of $(e^2\ln^2 q^2/p^2)^n$ terms. It is  supposed that $\mid q^2 \mid
\gg p^2_1,p^2_2$, i.e. the terms proportional to $(e^2 \ln q^2/p^2)^n$ are neglected.
Such approximation is called double logarithmic   approximation. If $p_1$ and $p_2$ are
on the mass shell, $p^2_1=p^2_2=m^2$, then the contribution of double logarithmic  terms
corresponds (at $q^2 < 0$)  to the electron electromagnetic  form factor, and to the
electron vertex function $\Gamma(p^2_1,p^2_2;q^2$) when $p^2_1$ and $p^2_2$ are off mass
shell.  The appearance of double logarithmic terms  can be most easily seen  by using the
Sudakov decomposition of the photon momentum $k$ of any photon line in diagrams of the
type of Fig.5:
\be k=u(p_1-\alpha_1 p_2)+ v(p_2 -\alpha_1 p_1) + k_{\perp},\label{31} \ee where
$k_{\perp}$ is the 4-vector orthogonal to $p_1$ and $p_2$:
\be k_{\perp} p_1 = k_{\perp} p_2 =0.\label{32} \ee
 The conditions
\be
 (p_1- \alpha_1 p_2)^2 = (p_2-\alpha_2 p_1)^2=0\label{33}
\ee are imposed on the parameters $\alpha_1$ and $\alpha_2$, which defines them
unambiguously. By considering the simplest diagrams of the type of Fig.5 -- the diagram
of Fig.4, where $G$ and $D$ are free propagators and $\Gamma_{\mu}=\gamma_{\mu}$, -- it
can be shown that the integration over $k_{\perp}$ does not result in the appearance of
logarithmic terms. The double logarithmic terms arise from the integration over $u$ and
$v$ and they are proportional to $\ln(q^2/p^2_2)$ and $\ln(q^2/p^2_1)$, respectively. The
consideration of more complicated diagrams can be performed in a similar way. In $n$-th
order of  perturbation theory the contributions of diagrams, which are different from one
another  by transposition of initial or final points of photon lines are equal. The final
result is:
\be \Gamma_{\mu} (p_1,p_2;q) = \gamma_{\mu} \exp \biggl [ -\frac{\alpha}{2\pi} \ln \biggl
( \frac{q^2}{p^2_1}\biggr ) \ln \biggl ( \frac{q^2}{p^2_2}\biggr )\biggr ], ~~~\alpha
=e^2=1/137 \label{34} \ee
 at $q^2 <0, p^2_1 < 0, p^2_2 <0$. The result is valid if
\be \ln \frac{q^2}{p^2_1} \gg 1,~~~\alpha \ln \frac{q^2}{p^2_1} \ll 1,~~~\alpha \ln^2
\frac{q^2}{p^2_1} \la 1\label{35} \ee In case of electrons on mass shell (electron form
factor) $p^2_1$ and $p^2_2$ must be replaced by $m^2$ and $q^2$ by $\mid q^2\mid$.

A similar technique was applied to calculate the double logarithmic asymptotics in many
other processes. But some specifics  may arise. For example, in case of high energy
electron elastic scattering in external electromagnetic field, where the emission of soft
photons by the electron must be accounted for, the corresponding formula for the cross
section is \cite{45},\cite{46}
\be d \sigma =d \sigma_0 \exp \biggl [ -\frac{2\alpha}{\pi} \ln \frac{\mid q^2 \mid}{m^2}
\ln \frac{E}{\Delta E}\biggr ], \label{36} \ee where $\sigma_0$ is the cross section in
Born approximation, $E$ is the electron energy and $\Delta E$ is the experimental
resolution, inside which the soft photons are not detected (or the uncertainty in the
measurement of electron final energy). Double logarithmic terms play an important role
in the production of narrow resonances $V$ in $e^+e^-$ annihilation,
\be
 e^+e^- \to V,\label{37}
\ee where the resonance can be produced not only when the $e^+e^-$ energy is equal to
$m_V$ (in this case the double logarithmic terms must be accounted for), but at a
slightly higher energy, when the $e^+$ and $e^-$ are emitting soft photons before
annihilation (see \cite{47} and references therein).

Last not least must be mentioned the book ``Quantum Electrodynamics'' written by
A.Akhiezer and V.B.Berestetsky \cite{48}. The first edition of the book was published in
1953 in Russian. Later the content of the book was extended and improved and new editions
of the book were published. The book was translated  into English. Now in my opinion it
is the best book on QED. I recommend to students  as well as to qualified specialists to
read it. (Of course, the parts of interest to them : to study the whole book is a very
hard task.) V.Berestetsky, a member of the ITEP Theory Group (in 1966-1977 the Head of
the ITEP Theory  Department) was also a co-author of one volume of the famous
Landau-Lifshitz Course of Theoretical Physics \cite{49}.

\section{Parity violation in weak interactions. The theory of $\beta$-decay and other weak
processes}

In 1955-56 the $\theta - \tau$ puzzle agitated all physicists. The  $K$-meson decays in 2
and 3 pions had been observed experimentally. Under the condition of parity conservation,
which was taken for granted at that time, one and the same kaon could not decay sometimes
into 2 and sometimes into 3 pions. For this reason most physicists  believed that
$\theta$, decaying into 2 pions and $\tau$, decaying into 3 pions, were different mesons.
As the precision of experiments grew, it became clear, however, that  their masses
coincided. At that time, in spring of 1956, Lee and Yang came up with their revolutionary
paper \cite{50}, in which they proposed the hypothesis of nonconservation of parity $(P)$
in weak interactions, that explained the $\theta-\tau$ puzzle. Moreover, Lee and Yang
calculated parity nonconservation effects in $\beta $ decay and in $\pi \to \mu\to e$
cascade decay.

Landau vigorously rejected the possibility of parity nonconservation, saying: ``space
cannot be asymmetric''. Pomeranchuk was attracted more by a hypothesis of
parity-degenerate doublets of strange  particles.
  A. Rudik and I decided to calculate some
additional  effects based on the assumption of parity nonconservation in weak
interactions, other than those considered by Lee and Yang. Our choice was the
$\beta-\gamma$ correlation. I made an estimate and found that the corresponding effect
had to be large. Rudic turned to a detailed calculation. After some  time he told me:

``Look, Boris, the effect vanishes. ''

``This cannot to be case'', I replied.\\
We started to examine the calculation and I noticed that Rudik, being a well-educated
theorist, had imposed the condition of charge $(C)$ invariance on the weak interaction
Lagrangian. As a result the coupling constants multiplying the parity non-conserving
terms turned out to be pure imaginary. The constants in the Lee and Yang paper were
arbitrary complex numbers. (If one assumes them to be pure imaginary, then all parity
non-conserving effects disappear.)

The question arose about  the connection between $C$ and $P$ invariance.  In a discussion
of this problem with Volodya Sudakov we remembered a paper by Pauli \cite{51} related to
this subject and published about a year ago. Although I had read this paper previously, I
had forgotten about it. In part the reason was that Landau regarded this paper with
scepticism  (although  he had a great respect for Pauli): he believed that the $CPT=1$
theorem ($T$-means the time inversion) was a trivial relation satisfied by any
Lagrangian, and no physical consequences could follow from the $CPT$ theorem. I noted
that Lee and Yang in their paper  did not mention the  $CPT$ theorem at all, and nothing
was said on the connection between $C$, $P$ and $T$ invariance. I read the Pauli paper
again, with more attention than at first,  and it became clear to me that if $P$ is
violated, then $C$ or $T$ or both had to be violated with certainty.

Then came the following idea. The existence  of two $K^0$-mesons with strongly different
lifetimes had been proved by Gell-Mann and Pais by assuming $C$-invariance of the weak
interaction Lagrangian. A similar proof can be made if $T$-invariance is applied. If $C$
and $T$-invariance are both violated to comparable degrees, then $K_S$  and $K_L$ would
have similar lifetimes. Experimentally, their lifetimes differ by two orders  of
magnitude. The conclusion is, that one of two conservation laws, $C$ or $T$ conservation,
had to be valid, at least approximately. Rudik and I considered a number of effects and
found, that $P$-odd  pair correlations of spin and momentum (the terms $\bosigma {\bf
p}$) appear if $C$ is violated and $T$ is conserved. In the opposite case they are
absent. (In my subsequent paper  \cite{52} -- paper V in the Appendix -- I proved  this
theorem  in a general form and found the type of $P$-odd terms corresponding  to
$T$-violation.) We wrote a paper and I told L.Okun about its content. Okun made the very
useful  remark that similar effects, which unambiguously differentiate the $C$ invariant
theories from $T$-invariant ones, appear in $K^0$ decays  too. We included this remark in
the paper  and I suggested to Okun to become a co-author. At first he refused, saying
that such a remark  deserves only to be mentioned in the acknowledgement, but later I
persuaded him. After that I reported our results to Pomeranchuk. Pomeranchuk decided that
we had to make our results known to Dau -- immediately, next Wednesday. On Wednesday
Dau's first reaction was to refuse to listen:
``I do not want to hear anything about parity nonconservation. This is nonsense !''\\
Chuk persuaded him:
``Dau, have patience for 15 minutes, listen to what the young people have to say.''\\
With heavy heart Dau agreed. I spoke not for long, perhaps for half an hour. Dau kept
silent, and then went away. Next day in the morning Pomeranchuk called me: Dau had solved
the parity nonconservation problem ! We were supposed to come to him immediately. By that
time both of Landau's papers -- on the conservation of combined parity
($CP$-conservation) and on the two-component neutrino  were ready with all formulations.
In the first paper he gave an acknowledgement  to us: ``I would like to express the deep
gratitude to B.Ioffe, L.Okun and A.Rudik, in a discussion  with whom the idea of this
work has arisen.''

Our paper \cite{53} and those of Landau \cite{54},\cite{55} were sent for publication
prior to the paper by Wu et al., \cite{56}, where the observation of electron asymmetry
in the decay of polarized nuclei was reported (i.e. the correlation between nuclear spin
and electron momentum). In this way the parity  nonconservation was discovered. Our
results implied that $C$-parity was not conserved in $\beta$-decay either. The
corresponding statement was added in proof to our paper. An similar statement  was made
in the paper by Wu et al \cite{56}, who referred to the paper by Lee, Oehme and Yang
\cite{57}, which in turn was published after our paper. In their Nobel lectures Lee and
Yang emphasized our priority in this problem. Later for this work we got the  USSR Prize
of Discovery.

Unfortunately, the story of the creation of these Landau works finished with an
unpleasant episode, which I shall mention for objectivity  reasons. A few days after the
presentation of his papers to ZhETF Landau gave an interview to a correspondent of
``Pravda'', the most important newspaper in the USSR. In this interview Landau told about
the problem of parity nonconservation and explained how he had solved it. The work of Lee
and Yang was not referred to  (not to mention our paper). All ITEP theoreticians were
indignant by this interview. Berestetsky  and Ter-Martirosyan visited Landau  and told
him, what they thought about this interview. The result followed immediately: both of
them were excommunicated from the Landau seminar. I did not give a piece of my mind
directly to Landau but presented it to his collaborators, who perhaps passed it on to
Landau. I was punished in another way: Landau cancelled my name in the acknowledgement in
his paper, leaving only Okun and Rudik. Here Pomeranchuk was not  be able to stick it any
longer. He visited Landau and said to him (this Chuk told me later): ``Boris explained to
you everything about $C$, $P$ and $T$. Without him your work would not have been done,
and you cancel his name in the acknowledgement !'' Landau made a compromise: he restored
my name in the acknowledgement, but not in  alphabetical order -- only as the second
name. Landau considered $CP$ conservation to be an exact law of the nature: he did not
admit its violation. Concerning $CP$ Landau would say exactly the same words on the space
asymmetry as he used to say previously with  regards to $P$ violation. I constructed  an
example of a Lagrangian in which $CP$ was violated and nothing happened to the vacuum and
tried to change Landau's mind, but he did not want to listen. The $CP$ violation was
discovered by Christenson, Cronin, Fitch and Turley \cite{58} in 1964, when Landau was in
very bad health, damaged in a car accident.

In a paper by Berestetsky et al. \cite{59}  the theory of $\beta$-decay was developed on
the most general grounds: all possible 5 variants of $\beta$-interaction with parity
violating terms were accounted for as well as the effects of Coulomb field and finite
nuclear size. $CP$-conservation was not assumed. Results were obtained for allowed
transitions, first order forbidden  transitions in intermediate and heavy nuclei (Coulomb
transitions), unique transitions $[~\Delta j=2$, (yes)] and  0-0 (yes) transition.
Angular distributions of electrons and neutrinos (equivalent to recoil nuclei
distributions) were calculated for unpolarized and polarized initial nuclei. The degree
of polarization of $\beta$-decay electrons was also determined. Particulary, it was shown
that $CP$-violation appears in the  distribution proportional to the term ${\bf n}_j[{\bf
n}_e,{\bf n}_{\nu}]$, where ${\bf n}_j$ is the direction of nuclear spin, ${\bf n}_e$ and
${\bf n}_{\nu}$ are the directions of the electron and neutrino momenta.

Okun and Pontecorvo demonstrated \cite{60}, that if a weak interaction with change of
strangeness by 2 units and with a coupling constant of the order of the Fermi coupling
constant exists, $G(\Delta S=2)\sim G_F$, then the mass difference $m_{K_1}-m_{K_2}$
would be greater than its actual value by many orders of magnitude. Therefore it was
proved that $G(\Delta S=2) \ll G_F$ and, probably, of the order of $G^2_F m^2$.

In 1958  Feynman and Gell-Mann \cite{61} and Marshak and Sudarshan  \cite{62} proposed a
universal form of $V-A$ weak interaction. According to \cite{61} the vector current that
enters the weak interaction belongs to the same isotopic triplet as the vector part of
electromagnetic current. (Earlier the same  hypothesis was formulated by Gershtein and
Zeldovich \cite{63}). This hypothesis was called the hypothesis of conserved vector
current -- $CVC$. As a check of the $CVC$ hypothesis Gell-Mann \cite{64} proposed the
measurement of the ``weak magnetism'' -- the correction term to the neutron-proton weak
vector form factor, which was calculated unambiguously in \cite{64} within the  framework
of $CVC$. Another method to check the $CVC$ hypothesis was suggested by Vaks and Ioffe
\cite{65}. Basing on  the $CVC$ theory they found the ratio of the $\pi \to e \nu \gamma$
decay width through the  vector current to the $\pi^0\to 2\gamma$ decay width and
predicted the electron spectrum in $\pi\to e\nu\gamma$ decay through  vector current.
Both checks were confirmed by experiment and now $CVC$ is one of the corner stones of the
Standard Model of weak interaction.


\section{Strong interactions and nuclear physics}

In the 1950-s K.Ter-Martirosyan produced two very good papers on nuclear physics. In the
first of them \cite{66} he calculated the excitation of nuclei by the electric Coulomb
field of a charged particle passing near the nucleus. In the second \cite{67} he and
Skornyakov solved the three-body problem in quantum mechanics for the case of an
infinitely small radius of two-body interaction.
 The solution of this problem has direct
applications to the case of interaction of three nuclei  at low energies, their
scattering, formation of bound states etc. In \cite{67} the equations for the three-body
problem were presented in differential form. Later they were reformulated by Faddeev in
integral form, which is more convenient for analysis. Till now these equations are widely
used in many branches of physics.

Great attention was paid to diffraction production of particles in $p-p$ and $p$-nucleus
collisions at high energies (see e.g. \cite{68}-\cite{70}). The main idea was that the
minimal momentum transfer from proton to nucleus in, say, pion production in $pA$
collisions is equal to $q=m\mu/E$, where $m$ and $\mu$ are the proton and pion masses,
and $E$ is the proton energy. If $1/q \gg R$ -- the nuclear radius, -- then the pion
production process proceeds outside the nucleus and the characteristics of such a process
can be calculated phenomenologically without the use of perturbation theory.

By this method a number of processes were calculated: elastic diffraction scattering in
$pp$ and $pA$ collisions, production of photons, mesons and meson pairs in $pA$
collisions, diffraction phenomena in deuteron-nucleus scattering, photon  production in
collisions  of mesons with nuclei etc. When Pomeranchuk reported the results of these
calculation at a seminar of the Lebedev Institute, Academician Skobelzyn asked: ``How can
it be that the production process proceeds outside the nucleus?'' Pomeranchuk explained
that  the wave function of the incoming particles overlaps with the shadow of the
nucleus, which results in a distortion of the wave function and gives rise to the
production processes. Then he continued his talk. After some time Skobelzyn repeated his
question. Pomeranchuk gave the same explanation, but in more detail. After another while
Skobelzyn repeated his question for a third time. Pomeranchuk's reply was: ``If you like,
you can consider this effect as immaculate conception''.

Pomeranchuk developed the  statistical approach to multiparticle  production in high
energy collisions, suggested by Fermi \cite{71}. Fermi supposed that in high energy
collisions the whole energy of the colliding particles (e.g. nuclei) is concentrated in
the small Lorentz contracted volume
\be \Omega  =\Omega_0 \frac{2M c^2}{W} =\frac{4\pi}{3} \biggl ( \frac{\hbar}{\mu c}
\biggr )^3 \frac{2Mc^2}{W},\label{37} \ee where $M$ and $\mu$ are the nuclear and pion
mass, respectively, and $W$ is the  total energy in the c.m. system. In this volume a
statistical equilibrium is set up and the probabilities of various states are
proportional to their statistical weights. Pomeranchuk remarked \cite{72} that in such a
small volume the produced particles are strongly interacting  with one another, which
results in the production of new particles and expansion of the fireball. This process
will continue till the distances between particles will be of the order of $\hbar/\mu c$.
It follows from Pomeranchuk's statement that the mean multiplicity $\overline{N}$ of
produced  particles at high  energies is proportional to $W$:
\be \overline{N} \approx \frac{W}{\varepsilon},\label{38} \ee where $\varepsilon$ is of
the order of a few hundred MeV, in comparison with Fermi's prediction $\tilde{N} \sim
\sqrt{W}$. The Pomeranchuk picture of multiparticle production was exploited by Landau
and Belenky \cite{73} in  their  hydrodynamical model of high  energy collisions and the
latter is widely used now in the description of heavy ion collisions.

When the charged pions had been discovered but not yet the $\pi^0$, the physical
community was divided in two groups: one of them believed that the $\pi^0$ exists and
will be found, the other denied its existence. Landau did not believe that the $\pi^0$
could exist and often repeated: ``God will not permit the existence of a diphoton''.
Later, when the $\pi^0$ was discovered, people who opposed his statements said in
discussions with him: ``But Dau, God did permit the existence of the diphoton''.

After the discovery of the $\pi^+$, $\pi^-$ and $\pi^0$ several papers at ITEP were the
devoted to determination of their spin and parity. The spin 0 of $\pi^0$ follows directly
from its decay into two $\gamma's$. It was natural to assume, that the $\pi^{\pm}$ spin
is also equal to 0. The negative parity of $\pi^-$ (and $\pi^+$) follows from the
observation of the reaction
\be \pi^- +d \to n+n \label{39} \ee for slow $\pi^-$ (captured from the $S$-level)
\cite{74}. In order to determine the parity of $\pi^0$  calculations  of the following
processes were performed \cite{75}:
\be
 \pi^- + d \to n + n +\gamma,\label{40}
 \ee
\be \pi^- + d \to n + n +\pi^0,\label{41} \ee when $\pi^-$ is captured from the $S$-state
in deuterium. The calculations were done in perturbation theory. (From the modern point
of view this is the correct approach, because the pion-nucleon interaction  tends to zero
at small momenta.) Two theoretical tricks were used in these calculations: 1) the
nonrelativistic expansion of the Hamiltonian in the operator form; 2) the use of the
completeness theorem for the determination of the total probability  in case of the
process (\ref{40}). Strong suppression of reaction (\ref{41})   in comparison with
(\ref{40}) was found in case of negative parity of $\pi^0$. Just  such suppression was
found experimentally. The conclusion was that the spin-parity of $\pi^{\pm,0}$ is
$J^p=0^-$. Later Pomeranchuk demonstrated that these processes can be calculated
phenomenologically, without the use of perturbation theory, leading to the same
conclusion \cite{76}.

Pomeranchuk had shown that at high energies the annihilation cross sections of
antiprotons with proton and neutron are equal \cite{77}, Okun and Pomeranchuk
demonstrated the equality at high energies of the total cross sections of $\pi^+$,
$\pi^-$ and $\pi^0$ on proton \cite{78}. Basing on the statement that the scattering
amplitudes at high angular momenta are determined by the exchange of the lightest
particle -- the pion, -- Okun and Pomeranchuk formulated the calculation method to
determine this amplitude \cite{79}.

The peak of the investigations of strong interactions in ITEP at the period under
consideration was the  famous Pomeranchuk theorem \cite{80}. This theorem states that
asymptotically at high energies the total cross sections of protons and antiprotons on
any target are equal. (The same statement holds for $n$ and $\overline{n}$ \cite{80}). In
the proof of this theorem Pomeranshuk exploited the dispersion relations for the
amplitudes of proton and antiproton scattering formulated in \cite{81},\cite{82} and the
assumptions that $\sigma(p)$, $\sigma(\overline{p})$ and the radius of interaction tend
to constant values at high energies. Although these assumptions are not correct according
to the modern point of view, the Pomeranchuk theorem is probably valid.

Finally,  I would like to mention the  book by A.Akhiezer and I.Ya.Pomeranchuk ``Some
problems in theory of nucleous'' \cite{83}. A lot of problems we considered in this book:
proton --neutron interaction, statistical description of heavy nucleous, the resonance
phenomena in the scattering of neutrons on nucleous, the neutron moderation in the
medium, the fission of heavy nucleous, the propagation of neutrons in crystalls.

\section{The contacts of Russian physicists with their Western colleques in 1955--1957 }

In the years 1945-1955 any contacts of Russian physicists with
their Western colleques were strictly forbidden. The same
prohibition refered also to publication of the papers in Western
journal, written by Russian authors. In this connection an
interesting episode happened.  At the year 1946 Pomeranchuk wrote
a paper about $\lambda$-limiting process in quantum field theory.
(the $\lambda$-limiting process for removing the infinities in
quantum field theories was suggested by Wentzel \cite{84} and
Dirac \cite{85}). Pomeranchuk demonstrated that this method is not
self-consistent and the results of its application are not unique.
The paper was send in Physical Review in 1946. At this time the
Cold War only starts and such violation of the rule was not
dangerous. For unknown to us reasons  the paper was published only
in 1949 \cite{86}, when the Cold war was in the full swing.
Pomeranchuk was afraid, that he will be punished. But nothing
happened, perhaps the censorship overlooked the appearance of this
paper.

The contacts were renewed in 1955. The first Western physicist,
who came to Moscow was G.K\"allen from Sweden. He presented a talk
at Lebedev Institute. Since it was a very unusual event a lot of
people came to listen his talk -- about 400 persons. In his talk
he presented the consideration of the Lee model at high energies
done by him and Pauli. The result was qualitative similar to the
result of Landau, Abrikosov and Khalatnikov -- the theory is
inconsistent at high energies -- the unphysical pole appears in
the expression for Green functions and effective charge. However,
Lee model has many drawbacks: the crossing symmetry was absent,
the causality principle was badly violated etc. These defects of
the model were mentioned by the participants of the seminar.

Some contacts of Russian and Western physicists took place at the
1-st International Conference on the Peaceful Use of Atomic
Energy, which took place at Geneve in 1955. The number of these
contacts was restricted: the atmosphere at the Conference was very
official. But in other aspect the Conference was very effective;
many papers from both sides were declassified and presented at the
Conference.

At 1956 very representative delegation of Western theoreticians
came to Moscow. Among the members of the delegation were: Dyson,
Gell-Mann, Marshak, Pais, Weisskopf, Low, Brueckner. A Conference
was organized at the Institute of Physical Problems (Kapitza
Institute) at which from the Russian side were presented the
talks: by Landau on the  inconsistency of QED (``zero charge''
problem); by Pomeranchuk about diffractional production and
scattering; Bogoljubov presented the strict proof of dispersion
relation for forward pion-nucleon scattering amplitude etc. The
Western participants also presented their talks. But the most
important for us, the Russian participants, were the private
conversation with leading Western scientists, in which we could
understand the direction of future development of our science.

After the publication of Landau, Abrokosov and Khalatnikov's paper
in approximately one year, Landau got a letter from Pauli. In this
letter Pauli informed him that his graduate student, Walter
Thirring, had found an example of the theory, in which there was
no zero charge problem -- the theory of scalar meson-nucleon
interaction. The manuscript of Thirring's paper was attached to
the letter. Landau gave this paper to Pomeranchuk and Pomeranchuk
asked me to check the paper. I studied Thirring's paper and came
to the conclusion that it was wrong. The origin of the mistake was
that, in the paper it was exploited the Ward identity arising from
differentiation  over nucleon mass, which in fact was violated by
renormalization. I told Pomeranchuk about this.

-- ``You should write a letter Pauli'' -- was Pomeranchuk
responce.

I hesitated: to write to Pauli, that his graduate student had make
a mistake and he, Pauli, had overlooked it ! Pomeranchuk insisted
and, finally, I wrote a letter. The answer I received, was not
from Pauli, but from Thirring. He accepted his error, and
Thirring's paper never was published. Later we became good
friends.

\section{Concluding remark}

Of course, only a part of the work done by ITEP theoreticians has been discussed in this
review. The selection of the presented papers is not free from subjectivity. This is
unavoidable. Only future historian, who can compare various points of view, will be able
to be objective. But the witness of all described events  has some advantage: he can
present to the reader the atmosphere in which the events proceeded. This will be
impossible for future historians.

I am very thankful to W.v.Schlippe, who improved the English and did valuable remarks. I
am indebted to M.N.Markina for preparation of the manuscript.

This work was supported in part by CRDF Cooperative Program grant RUP2-2961-MO-09 and
RFBR grant 12-02-00284-a.

\newpage

\addcontentsline{toc}{section}{{References }}

\newpage

\addcontentsline{toc}{section}{Appendix}

\centerline{\bf \large Appendix}

\bigskip

In the Appendix are presented few papers, done at ITEP, published in Russian at the years
1944-1957 and, perhaps, unknown  or almost unknown in the West.

\vspace{1cm}


\addcontentsline{toc}{subsection}{I ~D.D.Ivanenko and I.Ya.Pomeranchuk. On the maximal
energy reachable in betatron}

\begin{center}
{\large \bf  I ~On the maximal energy, reachable in betatron} \footnote{Doklady of AN
USSR, {\bf 44} 1944, 343. }

\bigskip
 {\it D.D.Ivanenko$^{1)}$ and I.Ya.Pomeranchuk$^{2)}$}

\bigskip

 $^{1)}$ Institute of Physics of Moscow State University

 $^{2)}$ Physico-Technical Institute of Academy of Science of USSR, Leningrad

\end{center}

\setcounter{equation}{0}
\def\theequation{\arabic{equation}}

\bigskip

Kerst \cite{Kerst1} was constructed a new  apparatus -- the betatron, with which  it was
possible to get the flux of electrons with the energy up to 10 MeV. The principle of
betatron action is: the electrons are accelerating by the electric field, arising from
the varying magnetic field inside the circle, where the electrons are moving. Unlike the
cyclotron, which can accelerate particles in nonrelativistic domain only, in betatron
there is no limits of energies, due to dependence of particle mass from their velocities.

However, there exist another circumstance, which set the limit of the energies  reachable
in betatron. The origin of such limit is the bremsstrahlung of electrons, moving on
circular orbit in the magnetic field. Indeed, the electron is accelerating, if it is
moving in the magnetic field, and in the accord of electrodynamics, it emits the energy.
It is easy to see, that because of large size of the orbit, the quantum effects plays no
role here. As it was shown in \cite{Pomeran1}, the electron moving in the magnetic field
$H$, is emitting the energy, equal to
\be
 -\frac{dE}{dx} =\frac{2}{3}  \biggl (
\frac{e^2}{mc^2}\biggr )^2 \biggl ( \frac{E}{mc^2}\biggr )^2 \biggl [\frac{{\bf
vH}}{c}\biggr ]^2 \label{1b}\ee on the unit of length. (Here $e$ and $m$ are the charge
and mass of electron, ${\bf v}$ and ${\bf E}$ are its velocity and energy. At the
derivation of (\ref{1b}) it was assumed, that $E \gg mc^2$.) In the Kerst's betatron on
the main part of the electron way $v\approx c$ and the direction of the velocity is
perpendicular to $H$. Therefore
\be -\frac{dE}{dx} =\frac{2}{3}r^2_0  \biggl ( \frac{EH}{mc^2}\biggr )^2, \biggl (
r_0=\frac{e^2}{mc^2}\biggr ). \label{2b}\ee The limiting energy $E_0$ is determined from
the condition, that (\ref{2b}) is equal to the energy gain of electron on the unit of
length \cite{Serber1}:
\be \frac{2}{3} r^2_0 \biggl ( \frac{E_0H}{mc^2}\biggr )^2 = \frac{e\biggl \vert
\frac{d\Phi}{dt}\biggr \vert}{2\pi R_0 c} =e R_0 \vert \dot{H}\vert \label{3b}\ee (Here
$R_0$ is the radius of the orbit, $\Phi$ -- is the induction flux, $\dot{H}=dH/dt$). It
follows from (\ref{3b})
\be E_0 = mc^2 \sqrt{\frac{2}{3} \frac{e}{c} \biggl \vert \frac{R_0\dot{H}}{r^2_0
H^2}\biggr \vert }.\label{4b}\ee

The magnitude of $H,\dot{H}$ and $E$, which are used now, give the  numerical value of
$E_0\approx 5\cdot 10^8$ eV. This number is only few times larger, than the value of
energy, $E\approx 10^8$ eV, which is proposed to get in the betatron under construction.

According to (\ref{4b}), $E_0$ is inversally proportional to the magnetic field strength
and is proportional to the square root of the energy gain on the unit of length.
Therefore, in order to increase $E_0$ it is necessary to go to smaller $H$ and larger
frequencies.

The radiation in the magnetic field also influence the bunch focusing, since the electron
energy is increasing more slowly with increasing of $H$ in comparison with the case, when
the radiation dumping is not accounted. This problem, however, requires  special
consideration.

\hfill Received 20.01.1944

\newpage

\addcontentsline{toc}{subsection}{II ~I.Ya.Pomeranchuk. To the theory of multiparticle
production in one event}
\begin{center}

{\large \bf II ~To the theory of multiparticle production in one event\footnote{Doklady
AN USSR {\bf 78}, 1951, 889.}}

\bigskip
{\it I.Ya.Pomeranchuk}

\bigskip

Academy of Science of USSR

\setcounter{equation}{0}
\def\theequation{\arabic{equation}}

\end{center}

\vspace{1cm}

As is known \cite{Wentzel2,Pauli2}, the nucleon-meson interaction, responsible for
nuclear forces, is very strong: $\pi$-mesons are demonstrating the strong interaction
with nucleons \cite{Brueckner2,Ashkin2}. As a consequence, the application of the meson
theory to the description of nucleon-meson interaction meets the enormous difficulties,
since the perturbation theory in not applicable here (may be with exception of low energy
domain). The nonapplicability of perturbation theory is especially evident at high
energies,  where the collisions with many particles production play the main role (the
showers production).

This conclusion directly  follows from the fact $\pi$-mesons (charged and neutral) are
pseudoscalar  particles \cite{Aamodt2,Tamor2}, which interaction energy with nucleons has
the form
$$U = g \overline{\psi}\gamma_5\psi\varphi + f\frac{\hbar}{\mu c} \varepsilon_{iklm}
\overline{\psi} \gamma_i\gamma_k \gamma_l \psi \frac{\partial \varphi}{\partial x_m}.$$
The constant $g^2/\hbar c$ is not small and under the action of the first term the
probability of the processes of the many particle production in one event is large, since
the ratio of the probability of $n$ particles  production to the ones of $n-1$ particles
is of the order of $g^2/\hbar c$. Under the action of term, proportional to $f$, this
statement becomes even stronger,  because the interaction increases with energy
\cite{Lewis2}. Therefore,  the consideration of nucleon-nucleon or nucleon-$\pi$-meson
collisions requires the methods, principially  different from perturbation theory.

Recently, Fermi \cite{Fermi2} paids attention to the possibly of using the methods of
statistical mechanics and thermodynamics to the description of nuclear collisions at high
energies. This possibility is a direct consequence of intensive interaction of colliding
and produced particles. According to Fermi the energy of colliding particles  is secreted
in a small volume, in which the statistical equilibrium is setted , where the
probabilities of various states are proportional to their statistical weights. By
choosing this volume equal to
\be \Omega = \Omega_0 \frac{2Mc^2}{W} = \frac{4\pi}{3} \biggl ( \frac{\hbar}{\mu c}
\biggr )^3 \frac{2Mc^2}{W}\label{1c}\ee ($M$ and $\mu$ are the nucleon (and pion) masses,
$W$ is the energy in c.m. system) Fermi calculated the probabilities of showers
production as a function of the number of produced particles and their masses at various
initial energies. (The total cross section Fermi assumed to be equal $\pi(\hbar/\mu
c)^2$.~\footnote{Perhaps, this value is to large. The cosmic rays data indicated that its
magnitude is equal to $(\hbar/\mu c)^2$.})

However, the value $\Omega(\ref{1c})$ looks not correct in the domain of relativistic
energies, where many particles are produced in the  collision.  Each produced particle
(nucleon, $\pi$-meson etc.) is strongly interacting with other particles. Therefore, if
$\Omega$ is chosen, as given by  (\ref{1c}), then in the domain of the size of order
$\hbar/\mu c$  many particles occur. All these particles   are strongly interacting with
one another and it is nonlegitimate  to use for the analysis the statistical  weights and
thermodynamical formulae of ideal gases, as it was done in \cite{Fermi2}. Instead of this
it is necessary to accept, that each particle is continuing the interaction with all
other, what results to the strongly increasing  of the number of produced particles. The
size of the whole system will strongly increasing. The expansion in all directions
proceeds with the velocity of light. This process will continue till the distances
between the particles will be of order $\sqrt{\sigma}\approx \sqrt{\hbar^2/\mu^2c^2}$.
After that the particles will move without interaction. At the final state the critical
density $n_0$
\be
n_0\approx \biggl ( \frac{\mu c}{\hbar}\biggr )^3\label{2c}\ee
could correspond to various particles number $N$, i.e. to the various volume
\be
\Omega \approx \frac{N}{n_0} \approx \biggl ( \frac{\hbar}{\mu c}\biggr )^3 N.\label{3c}\ee
Unlike to (\ref{1c}) $\Omega$ is not decreasing with the increase of energy, but even increasing
 with the number $N$ of particles in the shower. The probabilities $P(N)$ of various $N$
 can be obtained by using (\ref{3c}) and eq.(13) in \cite{Fermi2}:
 \be
 P(N) \approx \biggl [ \biggl ( \frac{4\pi}{3}\biggr)^{1/3} \frac{1}{\pi^{2/3}}
 \frac{W}{\mu c^2}\biggr ]^{3N} \frac{N^N}{(3N)!\biggl [\biggl ( \frac{N}{3}\biggr
 )!\biggr ]^3}.\label{4c}\ee
 An additional factor $N^N$ arises from (\ref{3c}) and the factor $[(N/3)!]^3$ accounts
 for nonindification of $\pi^-,\pi^+$ and $\pi^0$. In calculation of (\ref{4c}) it was
 supposed, that $N\gg 1$ and $N_{\pi^+} = N_{\pi^-} = N_{\pi^0}$. The nucleon pairs are
 not accounted  in (\ref{4c}) and all particles are considered as ultrarelativistic.

 The maximum of $P$ is reached at $\overline{N}$ equal
 \be
 \overline{N} \approx \frac{W}{\varepsilon} =
 \frac{\sqrt{2Mc^2W'}}{\varepsilon},\label{5c}\ee where $\varepsilon$ is of order of few
 hundred of MeV, $W'$ -- is the energy in the laboratory coordinate system, $W'\gg Mc^2$.
 Therefore, the mean number  of particles in the shower is increasing proportionally to
 $\sqrt{W'}$, but not $\sqrt[4]{W'}$, as it was obtained in \cite{Fermi2}.

 The equation (\ref{5}) is applicable in the case of central  collisions, when the whole
 energy of colliding particles is going to the produced ones. If the collision is not
 central, the number of produced particles will be much less, than given by (\ref{5}).

 The temperature of the system at the moment, when the critical density (\ref{2c}) is
 achieved, is of order $\varepsilon$, i.e. it is less, than $Mc^2$. Therefore, even at
 very high energies, the final temperature are smaller, then the temperatures needed for
 nucleon pair production. As a consequence in the star  observed in \cite{Lord2} it
 should not be nucleon pairs unlike the  conclusion done in \cite{Fermi2}.

Developing the method presented in \cite{Fermi2}, let us mention, that in nucleon-antinucleon
 collision  the dominating process is the nucleon-antinucleon annihilation, into pions.
  This is caused bye the fact, that the statistical weight of
 state, where the nucleon pair is absent is much larger, then the weight, where such pair
 persist. Independently of initial energy, the number of nucleon pairs in all showers
 shall be much smaller, than the number of $\pi$-mesons. The mean energy of particles in
 such showers (in c.m.s.) shall be independent of initial energy and to be order
 $\varepsilon$.

 In the showers, which are caused by noncentral collisions, the number of produced
 particles is much smaller, than $\overline{N}$. Such shower consist of the showers,
 developing  around each of colliding particles. However, even when $N \ll \overline{N}$,
 the mean energy of produced particles in each of the showers shall to be of the order of
 $\varepsilon$ in the coordinate system, where total moment of the shower is zero. In
 this system the distribution of particles is isotropic.

 I am indebted to L.D.Landau and Ya.B.Zeldovich for useful discussions.

 \vspace{1cm}

 \hfill Received 01.04.1954

\newpage

\addcontentsline{toc}{subsection}{III ~V.B.Berestetsky, B. L. Ioffe and I.Ya.Pomeranchuk.
 The heat conductivity of the completely
 ionized gas at high temperatures}
\begin{center}

{\large \bf III ~The heat conductivity of the completely ionized gas at high
temperatures} \footnote{This work was performed in ITEP in 1951 and was classified.
Declassified in 1972 and published in: I.Ya.Pomeranchuk, [Sobranie trudov], Nauka,
Moscow, 1972. }

\bigskip
{\it V.B.Berestetsky, B. L. Ioffe and I.Ya.Pomeranchuk}

\end{center}

\setcounter{equation}{0}
\def\theequation{\arabic{equation}}

We consider the heat conductivity of the completely ionized gas at the temperatures of
order 100 KeV. At such condition the velocities of ions are small in comparison with
electron velocities  and the contribution of ions to heat conductivity can be neglected.
The kinetic (Boltzmann) equation for the electron distribution function in the general
case, when persist the temperature gradient and the external electric field {\bf E} can
be written as:
\be (V \nabla n) - \frac{\partial n}{\partial{\bf p}} e {\bf E} = - J_{ie} -
J_{ee},\label{1a}\ee where $J_{ie}$ is the collisions integral of electrons with ions and
$J_{ee}$ is the integral of $ee$ collisions.
\be J_{ie} =\int v[~N ({\bf P,r}) n ({\bf p,r}) - N ({\bf P}',{\bf r})n({\bf p}',{\bf
r})~], \sigma_s (p,\theta') d \Omega d{\bf P},\label{2a} \ee
\be J_{ee} = \int [~n ({\bf P},{\bf r}) n({\bf p},{\bf r}) - n ({\bf P}', {\bf r}) n
({\bf p}',{\bf r})~] d w ({\bf p},{\bf p}',{\bf P}) d{\bf P},\label{3a}\ee Here $N({\bf
P,p})$ is the ion distribution function, $\sigma_s(p,\theta')d\Omega'$ is the
differential cross section of electron-ion collisions and $w$ is the probability of the
scattering at the $ee$ collisions.

As is known \cite{Landau}, at the kinetical  processes in gases in case of Coulomb
interaction the main role are playing the collisions with large impact parameters,
corresponding to small momentum transfers. For such collisions, i.e. for the collisions,
in which the momentum transfer $q=\mid {\borho}' - {\bf p} \mid \ll p$ is much smaller,
then the value of the momentum $p$, the cross section of electron scattering on the ion
is equal
\be \sigma_s(p,0) = 4\biggl ( \frac{e^2}{pv}\biggr )^2 \frac{1}{\theta^4} \label{4a}\ee
where $\theta$ is the scattering angle and the ion charge was putten  to 1. The
scattering probability at the collision of relativistic electrons can be found from the
scattering matrix element, which is calculated using the relativistic wave function with
account of retardation \cite{Heitler}:
\be H_{AE} = 4\pi c^2e^2 \left\{ \frac{(u^*_{01} u_1)(u^*_{02} u_2) -
(u^*_{01}\alpha_1u_1)(u^*_{02} \alpha_2 u_2)}{c^2({\bf p}_{01} - {\bf p}_1)^2 - (E_{01} -
E_1)^2} \right \},\label{5a}\ee In (\ref{5a}) $u$ -- are electron spinors, $p$ -- are the
momenta, $E$ -- are the electron energies, the indexes 1 and 2 refer to the first and
second electrons, the index 0 denotes the initial states. In the matrix element
(\ref{5a}) the exchange effects are not accounted and there is no antisymmetrization over
the final states. But in the collisions, where the impact parameters $d$ are much larger,
then the electron wave length $\lambda$, $d \gg \lambda$ , the quantum mechanical
exchange effect (i.e. the interference terms) can be disregarded. The classical  factor 2
will be accounted in the final formulas, where the integration will be performed over the
whole space instead of the half of space, as it should be done for identical particles.
At small momentum transfer we can put $u_{01}=u_1$ and $u_{02}=u_2$ and therefore
$$ u^*_{01} u_1 =1,~~~u^*_{01} {\bf \alpha} u_1 =\frac{{\bf v}_1}{c},$$
 \be u^*_{02} u_2 =1,~~~u^*_{02} {\bf \alpha} u_2 =\frac{{\bf v}_2}{c}.\label{6a}\ee
Let us expand in powers of
\be {\bf q} = {\bf p}_1 - {\bf p}_{01}\label{7a}\ee the energies difference in the
denominator of (\ref{5a})
\be E_1 - E_{01} = {\bf v}_1{\bf q}.\label{8a}\ee Then the probability of the scattering
of electron with momentum  ${\bf p_{01}}$ into the element $d{\bf p}_1$ of the momentum
space  is equal
\be dw =4e^4\frac{\biggl ( 1- \frac{{\bf v}_{01}{\bf v}_{02}}{c^2}\biggr )^2}{[q^2-({\bf
v}_{01} {\bf q})^2/c^2]^2} \delta ({\bf v}_{01} -{\bf  v}_{02}, {\bf q})d{\bf
p}_1.\label{9a}\ee (The argument of $\delta$-function, corresponding to energy
conservation, is expanded in powers of $q$ up to first order terms.)

As usually done by the heat conductivity calculation, let us suppose that the temperature
gradient as well the electric field are small. Therefore, the deviation of the electron
distribution from the equilibrium Maxwell ones is small
\be n=n_0(1+\chi),\label{10a}\ee Here $n_0$ is the Maxwell distribution of relativistic
electron gas \cite{Lifshitz}
\be n_0 dp =\rho_e \frac{\exp [-c\sqrt{m^2c^2 +p^2}/T]}{2\biggl (\frac{T}{mc^2}\biggr )^2
K_1 \biggl ( \frac{mc^2}{T}\Biggr ) +\frac{T}{mc^2} K_0 \biggl ( \frac{mc^2}{T} \biggr )}
\frac{d{\bf p}}{4\pi(mc)^3}, \label{11a}\ee where $\rho_e$ is the number of electrons in
cm$^3$, $T$-is the temperature in the units of energy, $K_0$ and $K_1$ -- are the
McDonald functions. The function $\chi_1$ describing the deviation from Maxwell
distribution shall have the from
\be \chi ={\bf p} \nabla T \chi_1(p) +{\bf p E} e\chi_2(p),\label{12a}\ee up to higher
order terms in $\nabla T$ and ${\bf E}$. $\chi_1(p)$ and $\chi_2(p)$ are the functions of
the  absolute values of the momentum. If the electron density is assumed to be constant,
the left hand side of eq.(\ref{1a}) can be represented in the form
\be \frac{\partial n_0}{\partial T} ({\bf v} \nabla T) -\frac{\partial n_0}{\partial p}
e\frac{\bf pE}{p}.\label{13a}\ee

Let us now determine the collision integrals. Since the equations are linear relative to
$\chi_1$ find first the equation for  $\chi_1$. The equation for $\chi_2$ will be found
in a similar way. Using (\ref{10a}) the integral of electron-ion collisions can be
written as
$$ J_{ie} =\int  v \{ N(P) n_0(p) [1+\chi ({\bf p,r})] - N (P') n_0 (p')[1+\chi({\bf
p}',{\bf r})]\} \times$$ \be \times \sigma_s(p,\theta') d \Omega' d{\bf P}=vn_0(p) \int
[\chi({\bf p},{\bf r})-\chi({\bf p}',{\bf r})] N(P)\sigma_s(p,\theta') d \Omega'd{\bf
P},\label{14a}\ee since in accord with energy conservation at the collision
\be N(P) n_0 (p) = N(P') n_0 (p').\label{15a}\ee The integration over ${\bf P}$ gives
\be \int N(P) d{\bf P} =\rho_i,\label{16a}\ee where $\rho_i$ is the number of ions in
cm$^3$. Using (\ref{12a}), we get
\be J_{ie} =\rho_i v n_0 (p) \nabla T\int ({\bf p}-{\bf p}')\chi_1 (p) \sigma_s
(p,\theta') d\Omega'.\label{17a}\ee The integral in (\ref{17a}) shall be proportional to
${\bf p}$. Multiplying the integral by ${\bf p}$ and substituting $\sigma_s$ from
(\ref{4a}) we find the expression for the collision integral of  electrons with ions
\be J_{ie} =4\pi L v \rho_i n_0 (p) \biggl ( \frac{e^2}{pv}\biggr )^2 ({\bf p} \nabla T)
\chi_1 (p).\label{18a}\ee In (\ref{18a})
\be L = \int\limits^{\theta_{\mbox{\footnotesize max}}}_{\theta_{\mbox{\footnotesize
min}}} \frac{d \theta}{\theta} =\ln \frac{r_{\mbox{\footnotesize
max}}}{r_{\mbox{\footnotesize min}}}\label{19a} \ee where $r_{max}$ and $r_{min}$ -- are
the maximal and minimal impact parameters. The maximal impact parameter is chosen
\cite{Landau} as the distance, where  the screening of the Coulomb field of the given ion
by neighbouring ions becomes important. In the absence of magnetic forces this distance
can be taken to be equal to the Debye-H\"uckel radius
\be r_{\mbox{\footnotesize min}} =\sqrt {\frac{T}{4\pi e^2} \frac{1}{\rho_i
+\rho_e}}.\label{20a}\ee The account of magnetic forces do not change the order of
magnitude of $r_{\footnotesize{\mbox{max}}}$ and, consequently, up to the small number of
order $1/L$, will not charge the value of the logarithm in (\ref{19a}). In the case of
high temperature of electrons the minimal impact parameter shall to be taken equal to
$\lambda =\hbar/p$, since at such distances the momentum transfers are not small.
Therefore
\be L =\frac{1}{2} \ln \biggl [ \frac{137}{8\pi} \biggl ( \frac{mc}{\hbar} \biggr )^3
\frac{1}{\rho_e} \biggl ( \frac{\rho}{mc}\biggr )^2 \frac{T}{mc^2}\biggr ].\label{21a}\ee

The account of only far separated collisions, which results to cut-off of logarithmically
divergent integrals at $\theta_{\footnotesize{\mbox{min}}}$ (or
$r_{\footnotesize{\mbox{max}}}$) restricts the accuracy of formulae
(\ref{18a}),(\ref{26a}) by the value of order $\sim 1/L \sim 0.1$ At the same order is
the  accuracy of the kinetical equations (\ref{43a}),(\ref{45a}) and the value of heat
conductivity , presented in Table 1.

For the calculation of collisions integral in case of $ee$ collision we use the general
expression, which was obtained in \cite{Landau} as a result of expansion at small
momentum transfer:
\be J_{ee}=\frac{\partial j_i}{\partial x_i},~~~ i = 1,2,3,\label{22a}\ee
\be j_i =\frac{1}{2}\int d{\bf P} \biggl [ n({\bf p}) \frac{dn({\bf P})}{\partial P_k} -
n({\bf P}) \frac{\partial n({\bf p})}{\partial p_k} \biggr ] q_i q_k dw,\label{23a}\ee In
(\ref{23a}) $dw$ is given by (\ref{9a}) (in the notation $v_{01}={\bf v},{\bf v}_{02}
={\bf V,p}_{01} ={\bf p},{\bf p}_1={\bf p}^{\prime}, {\bf q} ={\bf p}'-{\bf p})$. By
substitution of the expansion (\ref{10a}) for $n({\bf p})$ in (\ref{23a}) and by
accounting only the terms linear in $\chi$, we get for $j_i$
\be j_i =-\frac{1}{2} n_0 \int n_0 (P) \biggl [ \frac{\partial \chi ({\bf p})}{\partial
p_k} - \frac{\partial \chi ({\bf P})}{\partial P_k} \biggr ] q_i q_k dw d{\bf
P}\label{24a}\ee The equilibrium function $n_0$ do not contribute to $j_i$.

The substitution of $\partial \chi_1/\partial p_k$, given by (\ref{12a})
\be\frac{\partial \chi_1}{\partial p_k}= \chi_1(p) \frac{\partial T}{\partial x_k} +
({\bf P} \nabla T) \frac{p_k}{p}\frac{\partial \chi_1}{\partial p}\label{25a}\ee and
after transformation the resulting expression to the form convenient for angular
integration, it is obtained the formula for collision integral in case of $ee$ collisions
$$ J_{ee} =\frac{1}{2}({\bf p}\nabla T) \frac{1}{p^2} \left\{ \biggl [ n_0(\Phi_2 - 3\Phi_1) -
p\frac{\partial}{\partial p}(n_0(p)\Phi_1(p))\biggr ] \chi_1(p) -\right.$$
$$
-\biggl [ 4n_0(p) p\Phi_1(p) +p^2\frac{\partial}{\partial p} (n_0(p) \Phi_1(p))\biggr
]\frac{\partial \chi_1}{\partial p} -$$ $$ -  n_0(p) \Phi_1(p) p^2 \frac{\partial^2
\chi_1}{\partial p^2} - n_0 (p) [\overline{\Phi}_2(p)- 3\overline{\Phi}_1(p)] +$$ \be
\left.+ p\frac{\partial}{\partial p} [n_0(p) \overline{\Phi}_1(p)] +
3n_0\overline{\Phi}_3( p)+ p\frac{\partial}{\partial p}[n_0(p) \overline{\Phi}_3(p)] -
n_0(p) \overline{\Phi}^*_1(p)\right\},\label{26a}\ee where
\be \Phi_1(p) =\int n_0(P)\biggl ( \frac{\bf qp}{p}\biggr )^2 d{\bf P} dw, \label{27a}\ee
\be \Phi_2(p) =\int n_0(P) q^2 d{\bf P} dw,\label{28a}\ee
\be \overline{\Phi}_1(p) =\int n_0(p) \biggl ( \frac{\bf q p}{p}\biggr )^2 \chi_1(P)
d{\bf P} dw,\label{29a}\ee
\be \overline{\Phi}_2(p) =\int n_0(P) q^2 \chi_1(P) d{\bf P} dw,\label{30a}\ee
\be \overline{\Phi}_3(p) =\int \frac{\bf P p}{p}n_0(P) \frac{\partial \chi_1(P)}{\partial
P} \frac{v}{V}\biggl (\frac{\bf q p}{p}\biggr )^2 d{\bf P}  dw,\label{31a}\ee
\be \overline{\Phi}_1^*(p) =\int n_0(P) P  \frac{\partial \chi_1(P)}{\partial P} \biggl (
\frac{v}{V}\biggr )^2 \biggl (\frac{\bf q p}{p}\biggr )^2 d{\bf P} dw.\label{32a}\ee At
the derivation of (\ref{26a}) were accounted the equalities
\be {\bf V q} ={\bf v q}, ~\frac{\bf P q}{P} = \frac{v}{V} \frac{\bf p
q}{p}.\label{33a}\ee caused by the presence of $\delta$-function in (\ref{9a}). Let us
perform the integration in
$$ \Phi_1(p) = 4e^4\int n_0 (P) \biggl (\frac{\bf q p}{p}\biggr )^2
 \frac{\biggl ( 1-\frac{1}{c^2} {\bf v V}\biggr )^2}{\biggl [q^2-\frac{1}{c^2}({\bf vq})^2\biggr ]^2}
~ \delta  ({\bf v}-{\bf  V},{\bf q}) d {\bf q} d{\bf P}.\eqno{(27')}$$ First integrate
over the directions of the vector ${\bf P}$, after that over all values of ${\bf q}$. It
is convenient for the integration over the directions  of ${\bf P}$ to introduce the
coordinate system with the $z$-axe along ${\bf q}$. In this system the unit vectors ${\bf
P}/P$ and ${\bf p}/p$ can be represented as
\be \frac{\bf P}{P} = \{ \sin\beta \cos \delta;~~~\sin\beta \sin \delta;
~~~\cos\beta\},\label{34a}\ee
$$\frac{\bf p}{p} = \{ \sin\theta \cos \varphi;~~~\sin\theta \sin \varphi;
~~~\cos\theta\},\eqno{(34')}$$ and the integral takes the form
$$
\Phi_1(p) =4 e^4 \int\limits^{\infty}_0 n_0 (P) P^2dP \int \frac{dq}{q} \int
\frac{\cos^2\theta d \cos\theta d\varphi}{\biggl (1-\frac{\nu^2}{c^2} \cos^2\theta\biggr
)^2}\times $$ \be \times \biggl [ 1 -\frac{1}{c^2} V v (\cos \beta\cos\theta
+\sin\beta\sin\theta \cos (\varphi -\delta)\biggr ]^2 \delta(v \cos \theta - V \cos
\beta)d \cos \beta d\delta.\label{35a}\ee The integration over $\beta,\delta$ and
$\varphi$ gives
$$\Phi_1(p) =16\pi^2 e^4 \int\limits^{\infty}_0 n_0 (P) P^2 \frac{1}{V} dP \int
\frac{dq}{q} \int\limits^{\lambda}_0 \frac{\cos^2\theta d\cos \theta}{\biggl
(1-\frac{\nu^2}{c^2} \cos^2 \theta\biggr )^2} \times $$ \be \times \left \{ \biggl ( 1-
\frac{v^2}{c^2} \cos^2 \theta\biggr )^2 + \frac{V^2 v^2}{2c^2} \sin^2\theta
-\frac{v^4}{2c^4} \sin^2 \theta \cos^2\theta\right\},\label{36a}\ee where
\be \lambda = \left\{ \begin{array}{lll} \frac{v}{V} & \mbox{at} & v > V, \\ & &\\1 &
\mbox{at} & v < V. \end{array} \right. \label{37a}\ee Like the case of the collisions of
electrons with ions the integral over momentum transfer is equal
\be \int \frac{dq}{q} = \ln \frac{q_{\mbox{\footnotesize max}}}{q_{\mbox{\footnotesize
min}}} =\ln \frac{r_{\mbox{\footnotesize max}}}{r_{\mbox{\footnotesize
min}}}=L\label{38a}\ee Using the notation $\mbox{cos} \theta=u$ and $\beta=v/c$ we have
finally
$$ \Phi_1(p) =16 \pi^2 Le^4\int\limits^{\infty}_0 n_0 (P) \frac{P^2}{V} dP
\int\limits^{\lambda}_0 \frac{u^2 du}{(1-\beta^2 u^2)^2}\times $$ \be \times \left\{
(1-\beta^2 u^2)^2 +\beta^2 \frac{V^2}{2 c^2} (1-u^2) -\frac{1}{2} \beta^4
u^2(1-u^2)\right \}.\label{39a}\ee For $\Phi_2$ and $\Phi_3$ in a similar way we find
$$ \Phi_2(p) =16 \pi^2 Le^4\int\limits^{\infty}_0 n_0 (P) \frac{P^2}{V} dP
\int\limits^{\lambda}_0 \frac{du}{(1-\beta^2 u^2)^2}\times $$ \be \times \left\{
(1-\beta^2 u^2)^2 +\beta^2 \frac{V^2}{2 c^2} (1-u^2) -\frac{1}{2} \beta^4
u^2(1-u^2)\right \},\label{40a}\ee
$$ \overline{\Phi}_3(p) =16 \pi^2 Le^4\int\limits^{\infty}_0 n_0 (P) \frac{v^2}{V^3}
P^3 dP \frac{\partial \chi_1(P)}{\partial P} \int\limits^{\lambda}_{0}\frac{u^2
du}{(1-\beta^2 u^2)^2} \times
$$ \be \times \left\{u^2 (1-\beta^2 u^2)^2 -\frac{3}{2}(1-u^2)(1-\beta^2u^2)\biggl
( \frac{V^2}{c^2} -\beta^2 u^2\biggr ) +\frac{1}{2}(1-u^2)\biggl (
 \frac{V^2}{2 c^2}-\beta^2 u^2\biggr )\right \}.\label{41a}\ee
After substitution of (\ref{39a}),(\ref{40a}),(\ref{41a}) and the expressions for
$\Phi_1,\overline{\Phi}_2$ and $\overline{\Phi}_3$ in the collision integral, performing
the integration by parts, and  using the variable
\be x=\frac{cp}{T},\label{42a}\ee we get the following equation for the distribution
function $\chi_1(x)=4\pi(e^2/c)\rho_eL\chi_1(p)$:
\be 1 - \frac{3\alpha K_0(\alpha) + 6 K_1(\alpha) +\alpha^2 K_1(\alpha)}{[\alpha
K_0(\alpha) +2K_1(\alpha)] ~\sqrt{\alpha^2 +x^2}} =\hat{M} \chi_1(x),\label{43a}\ee where
$\alpha=mc^2/T$ and the linear integro-differential operator $\hat{M}$ is defined by
$$ \hat{M} \chi_1(x) =-\frac{\sqrt{\alpha^2+x^3}}{x^3} \chi_1(x) - \frac{1}{\alpha^2
K_0(\alpha) + 2\alpha K_1(\alpha)}~\frac{1}{x^2} \times$$ \be \times \left\{ A(x) \chi_1
(x) + B (x) \frac{d\chi_1(x)}{dx} +C(x) \frac{d^2 \chi_1(x)}{dx^2} +
\int\limits^{\infty}_0 K(x, y) \chi_1(y) dy\right\}.\label{44a}\ee The coefficients
$A(x),C(x)$ and the kernel $K(x,y)$ are given below
$$
A(x) =-\frac{x^2 \alpha^2}{\sqrt{x^2 +\alpha^2}} e^{-\sqrt{\alpha^2+x^2}} +\frac{1}{2}
e^{-\sqrt{\alpha^2+x^2}} \left\{ \sqrt{\alpha^2+x^2}(\alpha^2 +x^2 +\right.$$
$$\left.+2\sqrt{\alpha^2 +x^2}+2) ~\biggl [ 1-\frac{1}{2\beta} (1-\beta^2)\ln
\frac{1+\beta}{1-\beta}\biggr ] - \alpha^2\biggl ( 1-\frac{x}{2\beta}\biggr ) \biggl [
3-\frac{1}{2\beta} (3-\beta^2)\ln \frac{1+\beta}{1-\beta}\biggr ]\right\} +$$
$$ +\frac{1}{2} \int\limits^x_0 y \sqrt{\alpha^2 +y^2} e^{-\sqrt{\alpha^2+y^2}} dy
\left\{ -\frac{1}{\beta} \biggl [ \gamma (1-\beta^2) - \right.$$
$$-\frac{1}{2} (3-\beta^2)(1-\gamma^2) \ln \frac{1+\gamma}{1-\gamma}\biggr ]
+\frac{x}{2\beta^2}\biggl [ \gamma(5-3\beta^2) - $$
$$\left.-\frac{1}{2} (5-3\beta^2 -3\gamma^2 +\beta^2\gamma^2)\ln \frac{1+\gamma}{1-\gamma}
\biggr ]\right\};$$
$$ B(x) =-\frac{x}{4} e^{-\sqrt{\alpha^2+x^2}} \left\{ \frac{\alpha^2}{\beta^2} \biggl [
3+ 7\beta^2 -\frac{1}{2\beta} \biggl ( 3 +6\beta^2 -\beta^4 \ln \frac{1+\beta}{1-\beta}
\biggr ) \biggr ] - \right.$$
$$ -\frac{\alpha^2 x}{\beta} \biggl ( 3- \frac{3-\beta^2}{2\beta} \ln
\frac{1+\beta}{1-\beta} \biggr ) + 2 (\alpha^2 + x^2 +2\sqrt{\alpha^2 +x^2} +2)\times $$
$$ \left.\times \biggl [ \frac{1}{\beta^2} \biggl ( 1+3\beta^2 -\frac{1-\beta^4}{2\beta} ~\ln
\frac{1+\beta}{1-\beta}\biggr ) -\frac{x}{\beta} \biggl ( 1-\frac{1-\beta^2}{2\beta}
~\ln\frac{1+\beta}{1-\beta} \biggr) \biggr ] \right \} -$$
$$ -\frac{x}{4} \int\limits^x_0 y \sqrt{\alpha^2 +y^2} e^{-\sqrt{\alpha^2+y^2}} dy \left\{
\frac{1}{\beta^3}~\biggl [ \gamma(5-3\beta^4 +6\beta^2) -\right. $$
$$-\frac{1}{2} (5 + 6\beta^2 -3 \gamma^2 -3\beta^4 -6 \beta^2 \gamma^2 +\beta^4 \gamma^2)
\ln~ \frac{1+\gamma}{1-\gamma} \biggr ] -$$
$$ \left.- \frac{x}{\beta^2} \biggl [ \gamma(5- 3\beta^2) -\frac{1}{2} (5-3\beta^2 -3\gamma^2
+\beta^2 \gamma^2)\ln ~\frac{1+\gamma}{1-\gamma} \biggl ]\right \}$$
$$C(x) =- \frac{x^2}{4\beta^2} e^{-\sqrt{\alpha^2+x^2}}\left\{ 2 \biggl [ 1
-\frac{1}{2\beta} (1-\beta^2) \ln \frac{1+\beta}{1-\beta} \biggr ] (\alpha^2 + x^2
+2\sqrt{\alpha^2+x^2} +2) +\right.$$
$$ +\alpha^2 \biggl [ 3-\frac{1}{2\beta} (3-\beta^2)\ln \frac{1+\beta}{1-\beta} \biggr ]
- \frac{x^2}{4\beta^3} \int\limits^x_0 y\sqrt{\alpha^2 +y^2} e^{-\sqrt{\alpha^2+y^2}} dy
\times $$
$$\times \left \{ \gamma (5-3\beta^2) -\frac{1}{2} (5-3\beta^2 -3\gamma^2
+\beta^2\gamma^2) ~\ln \frac{1+\gamma}{1-\gamma}\right\};$$


%
%
 $$ K(x,y) = y\sqrt{\alpha^2+y^2}e^{-\sqrt{\alpha^2+y^2}} \left\{ \begin{array}{ll}
[(1-\gamma^2)(R_1 + R_2)+R_3 +R_4] & \mbox{at}~y > x,
 \\ & \\\nonumber
[(1-\beta^2)(\overline{R}_1 + \overline{R}_2)+\overline{R}_3+\overline{R}_4] &
\mbox{at}~y < x,
\end{array}\right. \eqno{~~}$$
where
$$ R_1=-1+\frac{1-\beta^2}{2\beta} \ln \frac{1+\beta}{1-\beta},$$
$$ R_2 =\frac{1}{2} \sqrt{\alpha^2+y^2} \biggl [ 5+(3\beta^2 -5)\frac{1}{2\beta} \ln
\frac{1+\beta}{1-\beta}\biggr ],$$
$$ R_3 =\frac{x}{2\beta}\biggl [ 5-4\beta^2 -3\gamma^2 +2\beta^2 \gamma^2 -\frac{1}{2\beta}
(1-\beta^2)(5-3\gamma^2)\ln \frac{1+\beta}{1-\beta}\biggr ],$$
$$R_4 =-\frac{x}{2\beta} \sqrt{\alpha^2+ y^2} \biggl [ 11-\frac{16}{3} \beta^2 -9\gamma^2
+4 \beta^2 \gamma^2 -$$
$$ -\frac{1}{2\beta} (11-9\beta^2- 9\gamma^2 + 7 \beta^2\gamma^2)\ln
\frac{1+\gamma}{1-\gamma}\biggr ],$$
$$\overline{R}_1 =\frac{1}{\beta} \biggl [ -\gamma +\frac{1}{2} (1-\gamma^2)\ln
\frac{1+\gamma}{1-\gamma}\biggr ],$$
$$\overline{R}_2 =\frac{x}{2\beta^2} \biggl [5\gamma -\frac{1}{2} (5-3\gamma^2)\ln \frac{1+\gamma}
{1-\gamma}\biggr ],$$
$$\overline{R}_3 =\frac{\sqrt{\alpha^2+y^2}}{2\beta} \biggl [ \gamma(5-3\beta^2
-4\gamma^2+2\beta^2\gamma^2) -\frac{1}{2}(1-\gamma^2)(5-3\beta^2) \ln
\frac{1+\gamma}{1-\gamma}\biggr ],$$
$$\overline{R}_4 =-x\frac{\sqrt{\alpha+y^2}}{4\beta^2} \biggl [ \gamma \biggl ( 11
-9\beta^2 -\frac{16}{3} \gamma^2 + 4\beta^2\gamma^2\biggr ) -$$
$$ -\frac{1}{2} (11-9\beta^2 -9\gamma^2 +7\beta^2\gamma^2)
\ln\frac{1+\gamma}{1-\gamma}\biggr ],$$ The notation
$$ \beta =\frac{v}{c} =\frac{x}{\sqrt{x^2+\alpha^2}};~~~~\gamma
=\frac{V}{c}=\frac{y}{\sqrt{y^2+\alpha^2}}.\eqno(44')$$ are used. The equation for the
function $\chi_2(x) = 4\pi(e^2/c)\rho_eL\chi_2(p)$ differs from (\ref{43a}) by the use of
unhomogeneous term in the left hand side of the second term in (\ref{14a}) and have the
form
\be \frac{1}{\sqrt{\alpha+x^2}} = \hat{M}\chi_2(x).\label{45a}\ee The border conditions
for the equations (\ref{43a}) and (\ref{45a}) are: $\chi=0$ and $\chi'(0)$ at $x=0$.
These conditions are obtained automatically  from the conditions of finiteness of
electron-ion collision integral. This  integral is dominating  at small electron
velocities, since the scattering cross section is inversely proportional to the 4-th
power of the relative velocity of colliding particles, tends to infinity in case of
electron-ion collisions and to finite limit in case of $e-e$ collisions. $\chi(x)$ is
proportional to $x^3$, since at small, $J_{ie}\sim (1/x^3)\chi(x)$. From this statement
follow the mention  above border conditions.

The knowledge of the distribution function allows to find easily the flux of heat and
heat conductivity. The flux of heat ${\bf j}$ is equal
\be {\bf j}=\int E_{\mbox{\footnotesize{kin}}} {\bf v} nd{\bf p} =\int
n_0E_{\mbox{\footnotesize{kin}}} {\bf v}\chi d{\bf p} =\int
n_0E_{\mbox{\footnotesize{kin}}} {\bf v}[{\bf p} \nabla T \chi_1(p) +e {\bf E
p}\chi_2(p)]d{\bf p},\label{46a}\ee since in case of equilibrium ${\bf j}$ vanishes. The
electric field ${\bf E}$, which enters eq.(\ref{46a}) is caused by redistribution of
charges in the gas. Such redistribution proceeds much before the equilibrium of
temperatures. The electric field ${\bf E}$ should be determined from the condition of the
vanishing of  the total current
\be {\bf J} =\int n{\bf v}d{\bf p} = \int n_0\chi{\bf v}d{\bf p} = \int n_0{\bf v} [({\bf
p} \nabla T)\chi_1(p) +e{\bf E p} \chi_2(p)]d{\bf p} =0.\label{47a}\ee Since the unique
direction in the problem in view is the direction of the gradient of the temperature, the
electric field and the current have the same direction. Choosing the axe $x$ for this
direction, we have from (\ref{47a})
\be eE_x \int n_0(p) v_x p_x \chi_2(p) d{\bf p} =- \frac{dT}{dx} \int n_0(p) v_x
p_x\chi_1(p) d{\bf p}.\label{48a}\ee It follows from (\ref{48a})
\be e{\bf E} = -\nabla T \frac{\int\limits^{\infty}_0 n_0(p) \frac{p^4}{E} \chi_1(p) dp}
{\int\limits^{\infty}_0 n_0(p) \frac{p^4}{E} \chi_2(p) dp}\equiv -\zeta \nabla
T.\label{49a}\ee By substituting of (\ref{49a}) into (\ref{46a}), we get
\be j_i =\int E_{\mbox{\footnotesize{kin}}} n_0(p) [\chi_1(p)-\zeta \chi_2(p)]
p_kv_id{\bf p} \frac{dT}{dx_k}.\label{50a}\ee From the other side
\be j_i =-\kappa\delta_{ik} \frac{\partial T}{\partial x_k},\label{51a}\ee where $\kappa$
is the heat conductivity. Therefore
$$
\kappa =-\frac{1}{3} \int E_{\mbox{\footnotesize{kin}}}n_0(p)[\chi_1(p) -\zeta \chi_2(p)]
{\bf pv} d{\bf p}=$$ \be =- \frac{4\pi}{3} \int\limits^{\infty}_0
E_{\mbox{\footnotesize{kin}}}[\chi_1(p) -\zeta \chi_2(p)]\frac{p^4 c^2}{E}
dp.\label{52a}\ee In terms of the variable $x$ and the functions $\chi_1(x),\chi_2(x)$
the final result looks like
$$ \kappa = -\frac{1}{12\pi} ~\frac{1}{L} \biggl ( \frac{T}{e^2}\biggr )^2 \biggl (
\frac{T}{m}\biggr )^{1/2}\frac{1}{\sqrt{\alpha}[2K_1(\alpha)+\alpha K_0(\alpha)]}\times$$
\be \times \int\limits^{\infty}_0 \biggl ( 1-\frac{\alpha}{\sqrt{\alpha^2+x^2}}\biggr )
[\chi_1(x) -\zeta\chi_2(x)] e^{-\sqrt{\alpha^2+x^2}} x^4 dx,\label{53a}\ee where
\be \zeta = \frac{\int\limits^{\infty}_0 e^{-\sqrt{\alpha^2+x^2}} \chi_1(x)
\frac{x^4}{\sqrt{\alpha^2+x^2}} dx}{\int\limits^{\infty}_0 e^{-\sqrt{\alpha^2+x^2}}
\chi_2(x) \frac{x^4}{\sqrt{\alpha^2+x^2}} dx}\label{54a}\ee The equation for distribution
function can be significantly simplified in two limits: nonrelativistic $(\alpha \gg 1)$
and ultrarelativistic $(\alpha \ll 1)$. In the nonrelativistic case it is convenient to
put
\be x' =\frac{1}{\sqrt{\alpha}} x =\frac{p}{\sqrt{mT}},\label{55a}\ee
$$ \chi' = \sqrt{\alpha \chi}.\eqno(55')$$
Being expressed in the terms of these variables the eq.'s(\ref{43a}) and (\ref{45a}) take
the form
$$ \frac{1}{2} (x^{\prime 2} -3) =- \frac{1}{x^{\prime 3} }\chi_1^{\prime} (x')
-\sqrt{\frac{2}{\pi}} \frac{1}{x^{\prime 2}} \left\{ A'(x') \chi'_1 (x') +B'(x')
\frac{d\chi'_1(x')}{dx'} +\right.$$ \be \left.+ C'(x') \frac{d^2\chi'_1(x')}{dx^{\prime
2}} + \int\limits^{\infty}_0 K'(x',y')\chi'_1 (y') dy'\right\}\label{56a}\ee
$$ 1=-\frac{1}{x^{\prime 3}} \chi^{\prime}_2(x') -\sqrt{\frac{2}{\pi}} \frac{1}{x^{\prime
2}} \left\{ A'(x') \chi^{\prime}_2 (x') +B'(x') \frac{d\chi^{\prime}_2(x')}{dx'} +
\right.$$ \be \left. + C'(x') \frac{d^2 \chi^{\prime}_2}{d x^{\prime 2}}
+\int\limits^{\infty}_0 K'(x', y')\chi_2(y') dy'\right\},\label{57a}\ee
\be A'(x) =-\frac{2}{3} x^2e^{-\frac{x^2}{2}} +\frac{\sqrt{2}}{x} \left\{ \Gamma \biggl (
\frac{3}{2}; \frac{x^2}{2}\biggr ) +\frac{2}{3} \Gamma \biggl (
\frac{5}{2};\frac{x^2}{2}\biggr )\right \}, \label{58a}\ee
\be B'(x) =-\frac{x}{3}(4-x^2)e^{-\frac{x^2}{2}} -\frac{2^{3/2}}{3} \frac{1-x^2}{x^2}
\Gamma \biggl ( \frac{5}{2};\frac{x^2}{2}\biggr ), \label{59a}\ee
\be C'(x) =-\frac{x^2}{3} e^{-\frac{x^2}{2}} -\frac{2^{3/2}}{3}\Gamma \biggl (
\frac{5}{2};\frac{x^2}{2}\biggr )\label{60a}\ee
\be K'(x,y) =\left\{ \begin{array}{ll} x^2 ye^{-y^{2/2}} \biggl ( \frac{1}{3}
-\frac{1}{5}
x^2 \biggr ) & \mbox{at}~~ y > x, \\& \\
\frac{y^4}{x} e^{-y^{2/2}} \biggl ( \frac{1}{3} -\frac{1}{5} y^2 \biggr ) & \mbox{at}~~ y
< x. \end{array}\right. \label{61a}\ee The expression for heat conductivity is also
simplified at $\alpha \gg 1$ (nonrelativistic case):
\be \kappa =-\frac{1}{12\pi} \frac{1}{\sqrt{2\pi}} \frac{1}{L} \biggl (
\frac{T}{e^2}\biggr )^2 \biggl ( \frac{T}{m}\biggr )^{1/2} \int\limits^{\infty}_0
x^{\prime 6} [\chi'_1(x') -\zeta\chi^{\prime}_2(x')] e^{-\frac{x^{\prime 2}}{2}}
dx',\label{62a}\ee where
\be \zeta = \frac{\int\limits^{\infty}_0 e^{-\frac{x^{\prime 2}}{2}} x^{\prime 4}
\chi^{\prime}_1(x') dx'}{\int\limits^{\infty}_0 e^{-\frac{x^{\prime 2}}{2}} x^{\prime 4}
\chi^{\prime}_2(x') dx'}.\label{63a}\ee The equations (\ref{43a}),(\ref{45a}) were solved
numerically for the temperatures: $T=50$ KeV ($\alpha =10.2)$, $T=100$ KeV $(\alpha
=5.1)$ and $T=200$ KeV ($\alpha=2.55)$. As the control of the method it was performed the
numerical solution of nonrelativistic equation (\ref{56a}),(\ref{57a}) and the solution
of equations for ultrarelativistic case by Chapman--Enskog method (see below). The
obtained values of $\kappa/(1/L)(T/e^2)^2(T/m)^{1/2}$ are shown in Table 1.

\bigskip
\hspace{11cm} Table 1

\bigskip
\begin{tabular}{c|c|c|c|c|c}  \hline &&&& \\
$T$ & $T << mc^2$ (non- & 50 KeV & 100 KeV & 200 KeV & $T \gg mc^2$ (ultra-\\
 & relativ.case) &&&& relativ.case) \\ \hline
 &&&&& \\ & 0.97 & 0.73 & 0.61 & 0.47 & $0.25\biggl (\frac{mc^2}{T}\biggr )^{1/2}$\\
 &&&&& \\ \hline
 \end{tabular}

\bigskip
The value of $L$, which enters into the expression for $\chi$ can be written according to
(\ref{21a}) as (at $\beta=\rho_i=4\cdot 10^{22} $ cm$^{-3}$)
\be L =10.8 -\frac{3}{2} \ln \frac{mc^2}{T} +\ln \overline{x},\label{64a}\ee where
$\overline{x}$ -- is some effective value of $x$, corresponding to the maximum of
integrand in (\ref{53a}). The magnitudes of $\overline{x}$ and the values of $L$
calculated by the use of (\ref{64a}) are presented in Table 2.

\bigskip

\hspace{9cm} Table 2

\bigskip

\begin{center}

\begin{tabular}{c|c|c|c} \hline &&& \\
~~~$T$ & ~~~50 KeV & ~~~100 KeV & ~~~200 KeV\\ &&& \\ \hline &&& \\
~~~$\overline{x}$ & ~~~10 & ~~~7.5 & ~~~5.8 \\ &&&  \\ ~~~$L$ & ~~~9.6 & ~~~10.3 & ~~~11.1 \\
&&& \\
\hline
\end{tabular}

\end{center}

\bigskip

In the paper \cite{Landshoff} the heat conductivity for nonrelativistic gas was
calculated by Chapman-Enskog method. The result was: $\kappa/(1/L)(T/e^2)^2(T/m)^{1/2} =
0.95$ \footnote{There is a misprint  in the eq.(\ref{62a}) of \cite{Landshoff}: it should
be $\sqrt{2\pi}$ instead of $\sqrt{2}$.}. The difference of this number with the value,
presented above is in the limit of the accuracy of both calculations (the accuracy of the
result, obtained in \cite{Landshoff}, is about 2\% -- see \cite{Landshoff1}). ] Note,
that the decrease of heat conductivity with increasing   of temperature can be expected
{\it a priori}, since $\chi \sim lvc_v$ ($l$ -- is free path length, which is
proportional to $E^2$, $c_v$ -- is the gas specific heat). This estimation results in
nonrelativistic case to $T$ dependence $\chi \sim T^{5/2}$ and in ultrarelativistic case
to $\chi\sim T^2$.

\bigskip
{\bf \large The calculation of heat conductivity in the ultrarelativistic case}

\bigskip

Let us consider the case $T \gg mc^2$, but suppose, that $T$ is not very large and the
movement of ions can be disregarded. In this case the kinetic equation can be obtained by
expansion in powers of small parameter $\alpha =mc^2/T$.
Performing such an expansion in (\ref{43a}) and (\ref{45a}) (evidently one can put
$\beta=\gamma=1)$ we get the equations for distribution functions $\chi_1(x)$ and
$\chi_2(x)$:
$$ 1 -\frac{3}{x} =-\frac{1}{x^2}\chi_1(x) -\frac{1}{2x^2}\left\{x\chi_1(x)-x(4-x)
\frac{d\chi_1}{dx} -x^2\frac{d^2 \chi_1}{dx^2}-\right.$$ \be \left.-\frac{x}{6}
\int\limits^{\infty}_0 y^3 e^{-y} \chi_1(y) dy\right\} \label{65a}\ee
$$ \frac{1}{x} = -\frac{1}{x^2} \chi_2(x)-\frac{1}{2x^2} \left\{ x\chi_2(x) - x(4-x)
\frac{d\chi_2}{dx} -x^2\frac{d^2\chi_2}{dx^2} -\right.$$ \be -\left.\frac{x}{6}
\int\limits^{\infty}_0 y^3 e^{-y} \chi_2(y)dy\right\},\label{66a}\ee The heat
conductivity is given by:
\be \kappa =-\frac{1}{24\pi} \frac{1}{L} \biggl ( \frac{T}{e^2}\biggr )^2
c\int\limits^{\infty}_0 [\chi_1(x) -\zeta \chi_2(x)] x^4 e^{-x} dx,\label{67a}\ee where
\be \zeta =\frac{\int\limits^{\infty}_0 e^{-x} x^3 \chi_1(x) dx}{\int\limits^{\infty}_0
e^{-x} x^3\chi_2 (x) dx}.\label{68a}\ee Expand $\chi_1(x)$ and $\chi_2(x)$ over the
generalized Laguerre polinomials of the 3-rd power
\be \chi_1(x) =\sum^{\infty}_{n=0} p_n L_n (x),\label{69a}\ee
\be \chi_2(x) =\sum^{\infty}_{n=0} q_nL_n(x),\label{70a}\ee which satisfy the
differential equation
\be x\frac{d^2 L_n(x)}{dx^2}+(4-x) \frac{d L_n(x)}{dx} + n L_n(x) =0\label{71a}\ee and
orthogonality conditions
\be \int\limits^{\infty}_0 x^3 e^{-x} L_n(x) L_m(x) dx =\delta_{mn}\cdot
m!(m+3)!\label{72a}\ee The substitution of (\ref{69a}) into (\ref{65a}) and the account
of (\ref{71a}) gives:
$$(3-x) x=\biggl (1+\frac{x}{2}\biggr )\sum^{\infty}_{n=0} p_nL_n(x) +\frac{x}{2}
\sum^{\infty}_{n=0} p_nL_n(x) -$$ \be -\frac{x}{12} \sum^{\infty}_{n=0}
p_n\int\limits^{\infty}_0 y^3 e^{-y} L_n(y)dy.\label{73a}\ee Multiplay (\ref{73a}) by
$x^3e^{-x} L_m(x)$ and integrate  over $x$ in the limits from zero to infinity. Then
using the orthogonality  condition and the recurrence  formula for Laguerre polinomials
of the 3-rd power
\be
 xL_n(x) =2(n+2) L_n -L_{n+1} -
n(n+3) L_{n-1},\label{74a}\ee we find the following infinite system of equations for
determination of the coefficients $p_n$:
$$ -48\delta_{n0} + 168 \delta_{n1} -240\delta_{n2} =- 2p^{\prime}_0(\delta_{n0} -\delta_{n1})
-\frac{1}{2} n^2 (n+3) p^{\prime}_{n-1} +$$ \be + (n^2 +3n +3) p^{\prime}_n -\frac{1}{2}
(n+2) p^{\prime}_{n+1}.\label{75a}\ee We introduced the notations $p^{\prime}_n =n!
(n+3)! p_n$. The equation for  determination of $q^{\prime}_n$ differs from (\ref{75a})
by its left hand side, which  in this case is equal
\be -24(\delta_{n0} -\delta_{n1}).\label{76a}\ee For the determination of the heat
conductivity it is enough to know the zeroth and firsth  coefficients of expansion,
because
\be \kappa=-\frac{1}{\pi} \frac{1}{L} \biggl ( \frac{T}{e^2}\biggr )^2 c[p_0 -p_1
-\zeta(q_0 - q_1)],\label{77a}\ee where
\be \zeta =\frac{p_0}{q_0}.\label{78a}\ee In order to get the desirable accuracy it is
enough to restrict  ourself by the account of the first 4 equations in (\ref{75a}) and
(\ref{76a}). (The calculation  with account of 5 equations demonstrate that the error is
less than 2\% in $\kappa$.)

Neglecting $p^{\prime}_m$ and $q^{\prime}_n$  with $n \geq 5$  and solving the remaining
system of equations, we determine
\be \begin{array}{llll} p^{\prime}_1 =24, & p^{\prime}_0 =-24 & q^{\prime}_0 =-19.7,\\
&&&\\
q^{\prime}_1 =4.32, &&&\\ &&& \\p_1=1 & p_0=-4, & q_0=-3.29 &
q_1=0.18.\end{array}\label{79a}\ee By substitution of these values into (\ref{77a}) and
(\ref{78a}) we get the final result -- the expression for the heat conductivity in the
ultrarelativistic case
\be x=0.25 \frac{1}{L} \biggl (\frac{T}{e^2}\biggr )^2 c.\label{80a}\ee

The authors are thankful to A.S.Kronrod for the numerical solution of the equations.

\newpage

\addcontentsline{toc}{subsection}{IV ~B.L.Ioffe, L.B.Okun and A.P.Rudik. To the problem
of parity nonconservation in weak interactions}
\begin{center}

{\large \bf IV ~To the problem of parity nonconservation in weak interactions
\footnote{ZhETF {\bf 32}, 1957, 396.}}

\bigskip
{\it B.L.Ioffe, L.B.Okun and A.P.Rudik}
\end{center}

\setcounter{equation}{0}
\def\theequation{\arabic{equation}}

\vspace{1cm}

One of   possible explanation of the puzzle, arising in $\theta$ and  $\tau$-decays
 of $K$-mesons \cite{Orear3} is the hypothesis of parity nonconservation in weak
 interactions. Lee and Yang \cite{Lee3} had shown, that  parity nonconservation could
 not be observed basing on  already existing experimental data (except, surely,
 $K$-mesons decays). They discussed various experiments, which could clear up the problem
 of parity  nonconservation  in weak interactions. However, Lee and  Yang do not require the
 invariance of weak interactions relative to time reversal or charge
 conjugation\footnote{The general consideration is performed by Pauli \cite{Pauli3}. It is
 important do note, that if  parity is not conserved,  the charge conjugation is not
 equivalent to time reversal, since, as was proved
 by Pauli on the basis of spin and statistic connection, the Lagrangian shall be invariant
  relative
 to the product of the transformations of charge conjugation and reflections of all four
 coordinates.}. If the parity is not conserving ($\theta$ and $\tau$ are the same
 particle), then the existence  of long living $K^0$-meson \cite{Lande3} can be
 explained by supposing the conservation of the  charge parity or the time reversal. The
 consideration of correlation experiments, performed by Lee and Yang in fact corresponds
 to supposition of conservation of time parity and violation of charge parity.

 If the space parity is not conserving, it is possible to determine experimentally, what
 parity  is conserving -- charge parity or time parity. In the case of time parity
 conservation the longliving (odd under time reversal) $K^0$-particle can decay in 3
 pions, which are in $S$-state (or $3\pi^0$), what is impossible in case of charge parity
 conservation.

 We show below, that the supposition of charge parity conservation results
 to quite different conclusions, than the ones obtained by Lee and Yang. As is known
 \cite{Pauli3,Tolhoek3}, the invariance of the Hamiltonian under the charge conjugation
 results  to definite phase relations among the coefficients at various interaction
 terms. Let us consider first the $\Lambda^0 \to p+\pi$ decay. If the parity conservation
 is not required, the interaction Hamiltonian for this process has the form (the terms
 without derivative are considered):
 \be
 H=g(\overline{\psi}_p
 \psi_{\Lambda})\varphi_{\pi}+iG(\overline{\psi}_p\gamma_5\psi_{\Lambda})\varphi_{\pi} +
 g^* (\overline{\psi}_{\lambda}\psi_p) \varphi^+_{\pi} + iG(\overline{\psi}_{\Lambda}
 \gamma_5\psi_p)\varphi^+_{\pi}.\label{1d}\ee
The   transformation of charge conjugation, as usually, is given by equalities
$$\varphi'(x)= \varphi^+(x); ~~~\psi'(x) =-\overline{\psi}(x)
C^{-1};~~~\overline{\psi}'(x) =C\psi(x),$$ where the matrix $C$ satisfy the conditions
$$ C^T =-C;~~~CC^+ =1; ~~~\gamma^T_{\mu} =-C\gamma_{\mu} C^{-1}.$$ The Feynman  notations
are used
$$\gamma_{\mu} =\{\beta\boalpha,\beta\};~~~\gamma_5=-i\gamma_1\gamma_2\gamma_3\gamma_4;
~~~\hat{a} = \gamma_{\mu}a_{\mu} =-\beta\boalpha {\bf a} +\beta a_4.$$ By applying to
(\ref{1d}) the charge
 conjugation transformation and by taken into account that $\gamma^T_5=C\gamma_5 C^{-1}$,
 it is easy to find, that  the invariance of the Hamiltonian can be achieved if the
 constants $g$ and $G$ are real \footnote{We suppose, that the spinors, belonging to
 different fields
 are anticommuting. The supposition about their commutativity would result to the
 appearance of
 nonessential common phase factor.}

 In the case of  parity nonconservation the pseudoscalar terms can contribute to the
 square of the transition matrix. In order to find out if such terms appear in
 $\Lambda^0$-decay, let us consider the decay of polarized $\Lambda^0$ at rest. Using
 interaction Hamiltonian given by (\ref{1d}) and calculating  the square of the matrix
 element, we find after taken the sum over the proton spin (the proton-pion interaction
 is not accounted)
 $$ \sum\mid M \mid^2 =\{g^2\overline{u}_{\Lambda}(\hat{p} +m) u_{\Lambda}
 -G^2\overline{u}_{\Lambda} \gamma_5(\hat{p} + m)\gamma_5u_{\Lambda} +$$
 \be
 + igG\overline{u}_{\Lambda}[\gamma_5 (\hat{p} + m) +(\hat{p} + m)\gamma_5]
 u_{\Lambda}\}/2E_p,\label{2d}\ee  where $\hat{p}$ is the momentum and $m$ is the proton
 mass. The pseudoscalar terms (proportional to $\bosigma {\bf p}$, where $\bosigma$ is
 $\Lambda^0$ spin) could arise evidently from the interference term only. However this
 term vanishes, since $\overline{u}_{\Lambda}\gamma_5 u_{\Lambda}=0$. In a similar way it
 is easy to check, that the terms, proportional to   $\bosigma {\bf p}$, vanish in the
 case, when persist two types of coupling -- the scalar and the vector. From the
 statement, presented above it follows, that in subsequental processes $\pi^-+p\to
 \Lambda^0+K^0$, $\Lambda^0 \to p+\pi^-$, considered by Lee and Yang, are absent the
 pseudoscalar terms (proportional to ${\bf p}_{\Lambda}{\bf p}_{\pi}{\bf p}_p$), since
 the role of the first strong interaction process reduces to the creation of polarized
 $\Lambda^0$-particle.  Therefore, it is impossible to conclude if the parity is
 conserved or not in this process by measuring of the proton angular distribution at
 $\Lambda^0$-decay. To distinguish the case of the parity conservation from the case of
 parity nonservation  is possible at radiative $\Lambda^0$-decay. In such  process the
 pseudoscalar terms are vanishing in the same way, as above, but the $\gamma$-quantum
 spectrum will be different, because in the case of parity nonconservation the both types
 of terms (proportional to $g^2$ and $G^2$) will contribute to the matrix element, what
 is impossible if parity is conserved.

 The other weak process, where the effects, connected with parity nonconservation could
 arise, is the $\beta$-decay. In the case of parity nonconservation the interaction
 Hamiltonian has the  form
 $$ H = (\overline{\psi}_p\psi_n)(c_S\overline{\psi}_e\psi_{\nu} +ic'_S\overline{\psi}_e
 \gamma_5\psi_{\nu}) +
 (\overline{\psi}_p[\gamma_{\mu},\gamma_{\nu}]\psi_n)\{c_T(\overline{\psi}_e[\gamma_{\mu},
 \gamma_{\nu}]
 \psi_{\nu}) +$$
 \be
 + ic'_T (\overline{\psi}_e\gamma_5 [\gamma_{\mu},\gamma_{\nu}]\psi_{\nu})\} + \mbox{Herm. conj.};
 ~~[\gamma_{\mu},\gamma_{\nu}] =1/2(\gamma_{\mu}\gamma_{\nu} -\gamma_{\nu}\gamma_{\mu}),\label{3d}\ee We restrict
 ourself by consideration of scalar and tensor interactions. The existing experiments
 indicate, that just these variants are realized. The use of charge conjugation
 invariance gives that the constants $c_S,c^{\prime}_S, c_T, c^{\prime}_T$ are real.

 Consider the $\beta$-decay of polarized nucleons. Calculate, using (\ref{3}), the square
 of the matrix element and perform the summation over the spins of neutrino. It is easy
 to convince yourself, that among the pseudoscalar terms remains only the terms, arising
 from the interference of the scalar and tensor  variants, while the terms, proportional
 to $c_Sc^{\prime}_S$ and $c_Tc^{\prime}_T$ are identically zero. After summing over
 electron spin we get ($\bf q$ is the neutrino momentum):
 $$(4/E_{\nu} E_e)(c^{\prime}_S c_T - c_S c^{\prime}_T) \left \{ \mbox{Im} \biggl (\int
 \overline{\psi}_p\psi_n \exp [-i({\bf p}_e -{\bf q}){\bf r}]d{\bf r} \times \right.$$
 $$ \times \int \overline{\psi}_n \bosigma \psi_p\exp [i({\bf p}_e -{\bf q}) {\bf r}]
 d{\bf r}\biggr ) \biggl ( {\bf q} E_e -{\bf p} E_{\nu}) + $$
 \be \left. + Re \biggl ( \int \overline{\psi}_p\psi_n \exp [-i({\bf p}_e -{\bf q}){\bf r}]
 d{\bf r} \int \overline{\psi}_n \boalpha \psi_p \exp [i({\bf p}_e -{\bf q}){\bf r}] d{\bf
 r}\biggr ) [{\bf p q}]\right\}. \label{4d}\ee The expression (\ref{4d}) can be easy
 calculated  in the case of polarized neutron $\beta$-decay and is equal to zero  in the
 case, when the proton polarization is not measured and the sum over its spin is
 performed. Expression (\ref{4d}) vanishes also in the case of allowed transitions.
 Indeed, at the allowed transitions only the first term in the figure brackets is
 accounted  in (\ref{4d}) and the $\exp[i({\bf p}_e-{\bf q}){\bf r}]$ can be replaced by 1.
 Then, because of the presence of scalar interaction the initial and final states
 correspond to the same value of the total momentum and its projection. But, as it is
 known, the diagonal matrix elements of the Hermitian  operator $\bosigma$ are real, what
 results to vanishing of (\ref{4d}). For the allowed transitions (\ref{4d}) is vanishing
 if it is accounted the interaction of the electron with the Coulomb field of nucleons
 (in the  first approximation in $Ze/\hbar v$).

 Therefore, the requirement of charge conjugation invariance, results to the absence of
 differences between the cases of parity conservation and nonconservation also in the
 simplest situations in $\beta$-decay.

 The authors are very indepted to Prof.V.B.Berestetsky and  V.V.Sudakov for discussion
 and to Prof.I.Ya.Pomeranchuk for  his interest to our work.

 \vspace{1cm}

 \hfill Received  by the editors  at 21.11.1956

 \vspace{1cm}

 Note added {\it in proof} (14.02.1957). Quite recently became known the results of
 experiments done by Prof. Wu on the $\beta$-decay of oriented nucleous $Co$. In these
 experiments was found the correlation of nuclear spin and direction of electron momentum
 (the term $\bosigma {\bf p}$). In accord with presented above statements this fact
 indicates on nonconservation of space and charge parities in $\beta$-decay.

\newpage

\addcontentsline{toc}{subsection}{V ~B.L.Ioffe. On the two possible schemes of parity
nonconservation}
\begin{center}

{\bf \large V ~On the two possible schemes of parity nonconservation }

\vspace{3mm} {\bf \large in weak interactions \footnote{ZhETF {\bf 32}, 1957, 1246.}}

\bigskip

{\it B.L.Ioffe}

\end{center}

\setcounter{equation}{0}
\def\theequation{\arabic{equation}}

\bigskip

One of the possible explanations of $K^+$ - mesons decay into two and tree $\pi$-mesons is the hypothesis
 of parity nonconservation in weak interactions \cite{Lee4}. If this hypothesis is
 accepted, then the question arises: if it is conserved the charge parity and the parity
 relative to time inversion in weak interaction. As is  known \cite{Pauli4}, the
 connection of spin with statistic requires the invariance of all interactions under the
 product of three transformations: the reflection of three space coordinate $I$, the time
 inversion $T$ and charge conjugation $C$, symbolically $ITC=1$. Therefore \cite{Ioffe4},
 in the case of parity violation in weak interactions $(I\not= 1)$ there are three
 possibilities : I) weak interactions are invariant under time reversal $(T=1)$, but not
 invariant under charge conjugation and $IC=1$; II) weak interactions  are invariant
 under charge conjugation $(C=1)$, but not  invariant under time reversal and $IT=1$;
 III)  weak interactions are noninvariant under charge conjugation as well as under time
 reversal, but $ITC=1$. It the last possibility is accepted, than the fact of the
 existence of long lived $K^0$ \cite{Lande4} looks as a pure accident, since the
 Gell-Mann and Pais arguments \cite{Gell-Mann4}, which are the base of this prediction
 holds only in the case of conservation of charge  parity or the parity under time
 reversal. This fact allows to throw away the last possibility and to consider the first
 two only.

 In this paper we consider, what physical phenomena can take place in each of these
 alternative possibilities.  The first of these possibilities, as was mentioned by
 Landau\cite{Landau4}, physically corresponds to the supposition, that all interactions
 are invariant under simultaneous change of left to right and to transfer from particles
 to antiparticles.  The physical sense of the second supposition is that all interactions
 are invariant if simultaneously  with the transfer from left to right proceeds the
 backward  movement in time.

 Consider first the scheme I, when at the parity violation is conserved the time
 inversion. Let at $t \to -\infty$  exist the set of particles in the state $a$, the
 momenta in this state are ${\bf p}_a$ and the mean values of spins are ${\bf s}_a$. As a
 result of interaction this system of particles at $t\to +\infty$ transfers into the
 other system with momenta ${\bf p}_b$ and mean values of spins ${\bf s}_b$. From the
 time reversal invariance than follows \cite{Blatt4} that the matrix element of the
 transition $S^I_{ab}({\bf p}_a,{\bf s}_a;{\bf p}_b,{\bf s}_b)$ is connected with
 the matrix element of the inverse transition $S^I_{ba} ({\bf p}_b,{\bf s}_b; {\bf
 p}_a,{\bf s}_a)$ by
 \be
 S^I_{ab} ({\bf p}_a, {\bf s}_a; {\bf p}_b, {\bf s}_b) = S^I_{ba}(-{\bf p}_b -{\bf s}_b;
 -{\bf p}_a, -{\bf s}_a).\label{1f}\ee The matrix element $S_{ba}$, considered as a
 function of its arguments, do not have, generally, the same functional form as the
 function $S_{ab}$. This circumstance do not allow to have a direct profit from the
 relation (\ref{1f}). However, if the transition $a\to b$ proceeds due to weak
 interaction, then in the first nonvanishing approximation in this interaction the
 detailed   balance relation takes place
 \be
 S_{ab} ({\bf p}_a, {\bf s}_a; {\bf p}_b, {\bf s}_b) = -S^*_{ba}({\bf p}_b, {\bf s}_b;
 {\bf p}_a, {\bf s}_a).\label{2f}\ee (The relation (\ref{2f}) is legitimate if transition
 proceeds because of weak interaction, but it is not required that the movement of the
 particle is described by free wave functions.) By excluding  $S_{ba}$ from (\ref{1f}) and
 (\ref{2f})  we
 find that in the case of the invariance under time reversal the transition matrix
 element satisfies  the equality
 \be
 S^I_{ab} ({\bf p}_a, {\bf s}_a; {\bf p}_b, {\bf s}_b) =- S^{I*}_{ab} (-{\bf p}_a, -{\bf
 s}_a; -{\bf p}_b, -{\bf s}_b)\label{3f}\ee and. consequently, the probability of the
 transition satisfies the relation
 $$ W^I_{ab} ({\bf p}_a, {\bf s}_a; {\bf p}_b, {\bf s}_b) = W^I_{ab} (-{\bf p}_a, -{\bf
 s}_a; -{\bf p}_b, -{\bf s}_b). \eqno(\mbox{A})$$

 The analogous consideration can be performed in the case of the second possible scheme,
 when all interactions are invariant under charge conjugation. In this case instead  of
 time inversal transformation it is necessary to consider the inversion of all four
 coordinates. The matrix elements of the direct and inversed transitions are connected by
 the relation
 \be
 S^{II}_{ab} ({\bf p}_a, {\bf s}_a; {\bf p}_b, {\bf s}_b) = S^{II}_{ba} ({\bf p}_b, -{\bf
 s}_b; {\bf p}_a, -{\bf s}_a).\label{4f}\ee The equality (\ref{2f}), which is based on
 the Hermitancy of Hamiltonian is valid in this case also. The substitution of (\ref{2f})
 into (\ref{4f}) gives
 \be
 S^{II}_{ab} ({\bf p}_a, {\bf s}_a; {\bf p}_b, {\bf s}_b) =-S^{II*}_{ab} ({\bf p}_a,
 -{\bf s}_a; {\bf p}_b, -{\bf s}_b).\label{5f}\ee Therefore, in the case of the scheme
 II the transition probability satisfy the relation
 $$W^{II}_{ab} ({\bf p}_a, {\bf s}_a; {\bf p}_b, {\bf s}_b) = W^{II}_{ab} ({\bf p}_a,
 -{\bf s}_a; {\bf p}_b, - {\bf s}_b). \eqno(\mbox{B})$$ By the use (A) and (B) it is easy
 to determine the general form of the transition probability for both schemes of parity
 nonconservation.

 Consider, for example, the decay of polarized $\Lambda$-particle at rest: $\Lambda^0 \to
 p +\pi^-$. In general, $\Lambda$ decay is characterized by three vectors: the $\Lambda$
 spin ${\bf s}_{\Lambda}$, proton momentum ${\bf p}_p$ and proton spin ${\bf s}_{\beta}$.
 We are interesting by pseudoscalar quantities, arising from parity nonconservbation.
 From three vectors it is possible to constract three such expressions ${\bf s}_p{\bf
 p}_p$, ${\bf s}_{\Lambda}{\bf p}_p$, $[{\bf s}_{\Lambda}{\bf s}_p]{\bf p}_p$. From (A)
 and (B) it follows that in the case of scheme $I$ the probability can contain the terms ${\bf
 s}_p{\bf p}_p$  and ${\bf s}_{\Lambda} {\bf p}_p$, whereas in the case of the scheme
 $II$  in the probability only the terms,  proportional to $[{\bf s}_{\Lambda}
  {\bf s}_p]{\bf
 p}_p$ can appear. So, in the scheme I at $\Lambda$ decay one can expect the proton
 polarization along (or opposite) the direction of proton momentum, or if
 $\Lambda$-particle is polarized the protons will be mainly directed along (or opposite)
 the $\Lambda$ spin. (This effect was considered by Lee and Yang\cite{Lee4}.) In the
 scheme II the effect of parity nonconservation can be found only by observation of
 proton polarization at the decay of polarized $\Lambda$. Or, what is equivalent, by the
 measurement of the directions of proton spin and momentum relative to the normal to the
 plane of $\Lambda$ production. In a similar way it is possible to find the pseudoscalar
 terms allowed in other weak decays for each scheme. The results are presented in the
 Table. (The indexes $\Lambda, p, e,\mu$ are related  to the spins and momenta of
 $\Lambda$, proton, electron and $\mu$-meson correspondingly; in the case of
 $\beta$-decay: ${\bf I}_N$ -- is the spin of initial nucleous, ${\bf p}_N$ -- is the
 momentum of recoil  nucleous, the averaging is performed over all other momenta.)

 \newpage

 \hspace{9cm} {\bf Table}

\bigskip

 \begin{center}

 \begin{tabular}{c|c|c} \hline

The type & ~~Scheme I & ~~Scheme II \\
of decay & ~~$T=inv$  & ~~$C=inv$ \\ \hline $\Lambda$ & ~~${\bf s}_p{\bf p}_p; ~{\bf
s}_{\Lambda}{\bf p}_p$ & ~~$[{\bf s}_{\Lambda}{\bf s}_p]{\bf p}_p$\\ && \\ $\beta$ &
~~${\bf s}_e{\bf p}_e;~ {\bf s}_e{\bf p}_N$ & ~~$[{\bf s}_e {\bf I}_N]{\bf p}_e$ \\ &
~~${\bf I}_N{\bf p}_e; ~{\bf I}_N{\bf p}_N$ & ~~$[s_e{\bf I}_N]{\bf p}_N$\\ && \\$\pi$ &
~~~${\bf s}_{\mu}{\bf p}_{\mu}$ & ~~~~-- \\ && \\$\mu$ & ~~${\bf s}_{\mu}{\bf p}_e; ~{\bf
s}_e{\bf p}_e$ & ~~$[{\bf s}_{\mu}{\bf s}_e]{\bf p}_e$
\\ \hline

\end{tabular}
\end{center}

\bigskip

The nucleous dipole momenta are absent in the scheme I \cite{Landau4}, but could persist in the
 scheme II. Indeed, the energy of interaction of the dipole momentum with electric field
 is proportional to ${\bf sE}$. At the time reflection ${\bf s} \to -{\bf s}$ and ${\bf
 E} \to {\bf E}$, at the reflection of all four coordinates: ${\bf s} \to -{\bf s}$ and
 ${\bf E}\to -{\bf E}$. Therefore, ${\bf sE} \to -{\bf sE}$ in the scheme I and  ${\bf
 sE} \to {\bf sE}$ in the scheme II.

 The author is thankful to L.B.Okun and A.P.Rudik for very useful discussions as well to
 L.D.Landau for encouragement.

 \vspace{1cm}

 \hfill Received by the editors at 19.01.1957

\end{document}